\documentclass[iop]{emulateapj}
\usepackage{apjfonts}
\usepackage{graphicx}
\usepackage{color}

\usepackage{epstopdf}
\usepackage{amsmath}

\def\ltsima{$\; \buildrel < \over \sim \;$}
\def\simlt{\lower.5ex\hbox{\ltsima}}
\def\gtsima{$\; \buildrel > \over \sim \;$}
\def\simgt{\lower.5ex\hbox{\gtsima}}

\newcommand\lsim{\mathrel{\rlap{\lower4pt\hbox{\hskip1pt$\sim$}}
\raise1pt\hbox{$<$}}}
\newcommand\gsim{\mathrel{\rlap{\lower4pt\hbox{\hskip1pt$\sim$}}
\raise1pt\hbox{$>$}}}

\allowdisplaybreaks
\usepackage{color}

\shorttitle{Small kicks, Large outcomes }
\shortauthors{Naoz et al. }
\begin{document}

\title{  The Combined Effects of Two-Body Relaxation Processes and the Eccentric Kozai-Lidov Mechanism on the EMRI Rate}

\author{  Smadar Naoz$^{1,2}$, Sanaea C. Rose$^{1,2}$,  Erez Michaely$^{1,2}$, Denyz Melchor$^{1,2}$, Enrico Ramirez-Ruiz$^3$, Brenna Mockler$^3$, Jeremy D. Schnittman$^{4,5}$}

\altaffiltext{1}{Department of Physics and Astronomy, University of California, Los Angeles, CA 90095, USA}
\altaffiltext{2}{Mani L. Bhaumik Institute for Theoretical Physics, Department of Physics and Astronomy, UCLA, Los Angeles, CA 90095, USA}
\altaffiltext{3}{Department of Astronomy and Astrophysics,  University of California,Santa Cruz, CA 95064, USA}
\altaffiltext{4}{NASA Goddard Space Flight Center, Greenbelt, MD 20771, USA}
\altaffiltext{5}{Maryland Joint Space-Science Institute, College Park, MD 20742, USA}
\begin{abstract}
 Gravitational wave (GW) emissions from extreme-mass-ratio inspirals (EMRIs) are promising sources for low-frequency GW-detectors. They result from a compact object, such as a stellar-mass black-hole (BH), captured by a  supermassive black hole (SMBH). Several physical processes have been proposed to form EMRIs. In particular, weak two-body interactions over a long time scale (i.e.,  relaxation processes) have been proposed as a likely mechanism to drive the BH orbit to high eccentricity. Consequently, it is captured by the SMBH and becomes an EMRI.  Here we demonstrate that EMRIs are naturally formed in SMBH binaries. Gravitational perturbations from an SMBH companion, known as the eccentric Kozai-Lidov (EKL) mechanism, combined with relaxation processes, yield a significantly more enhanced rate than any of these processes operating alone. Since EKL is sensitive to the orbital configuration, two-body relaxation can alter the orbital parameters, rendering the system in a more EKL-favorable regime. As SMBH binaries are expected to be prevalent in the Universe, this process predicts a substantially high EMRI rate. 
\end{abstract}
\maketitle

\section{Introduction}

Extreme-mass-ratio inspirals (EMRIs) arise from the capture of a stellar mass compact object by a supermassive black hole (SMBH). The  Gravitational Wave (GW) emission from such a system is expected to be at the mHz band, thus a promising signal for the Laser Interferometer Space Antenna (LISA), as well as other mHz detectors, such as TianQin \citep[e.g.,][]{Amaro-Seoane+17,Robson+19,Baker+19,Mei+20}. Thus, the rate estimation of EMRIs is of high importance for these detections. 

EMRI rate estimation studies often focused on the "loss cone" mechanism, in which stellar mass black holes (BHs) undergo weak two-body scatterings and over time are able to reach high eccentricities \citep[e.g.,][]{Hopman+05,Aharon+16,Amaro18,Sari+19}. Additionally, weak two-body interactions can also lead to mass segregation if the BH is more massive than the surrounding population of stars \citep[e.g.,][]{Hopman+06seg,Alexander+09,Preto+10,Amaro+11,chen+18}.  Other physical processes have also been suggested to contribute to the formation of EMRIs, for example, the tidal separation of BH binaries by SMBHs was suggested to form a low eccentricity LISA event \citep[e.g.,][the latter also include the effects of mass segregation]{Miller+05,Raveh+21}. Furthermore, accretion disks around SMBHs in active galactic nuclei (AGN) have been suggested to further increase the EMRI rate \citep[e.g.,][]{Pan+21}.

Of particular interest here is the formation of EMRIs in SMBH binaries. Thanks to the hierarchical nature of galaxy formation, and since almost every galaxy hosts a SMBH at its center, SMBH binaries are expected to be a common phenomenon \citep[e.g.,][]{DiMatteo+05,Hopkins+06,Robertson+06,Callegari+09,Li+20Pair}. Observations of AGN pairs, which are typically few kpc (and more) apart, suggest that these configurations may lead to the formation of SMBH binaries with parsec to sub-parsec separations \citep[e.g.,][]{Komossa+03,Bianchi+08,Comerford+09bin,Comerford+18,Green+10,Liu+10kpc,Smith+10,Stemo+20}. Moreover, observations of SMBH binaries on wide orbits, as well as some sub-parsec candidates, seem to support this idea   \citep[e.g.,][]{Sillanpaa+88,Rodriguez+06,Komossa+08,Bogdanovic+09,Boroson+09,Dotti+09,Batcheldor+10,Deane+14,Runnoe+17,Pesce+18}. Lastly, a combination of theoretical and observational studies suggested that our own galactic center may also host a companion (albeit a small one)  
\citep[e.g.,][]{Hansen+03,Maillard+04,Grkan+20,Gualandris+09,Chen+13,Generozov+20,Fragione+20,Zheng+20,Naoz+20,Gravity+20}.

A SMBH companion gravitationally perturbs the orbit of a stellar-mass BH via the Eccentric Kozai-Lidov mechanism \citep[EKL, e.g.,][see latter for review]{Kozai,Lidov,Naoz16}.  These perturbations can result in extreme eccentricities \citep[e.g.,][]{Li+13,Li+14,NaozSilk14}, which can lead to the formation of EMRIs \citep[e.g.,][]{Bode+14,Haster+16}.  A similar process is often considered for the production of tidal disruption events  \citep[e.g.,][Mockler et al. in prep.]{Chen+08,Chen+09,Chen+11,Chen+13,Li+15}.

The EKL approach often neglects collective dynamical interaction because these interactions operate on much longer timescales. In particular, relaxation by gravitational encounters typically takes place on such long time scales, compared to other physical processes (see below), and thus often is neglected when considering EKL processes. Here we show that the combined effect of EKL and relaxation processes enhances the EMRI formation efficiency more than any of these processes operating alone. Furthermore, the two-body relaxation processes overcome the general relativistic precession that suppresses EKL resonances. We begin by describing the methodology of the system in Section \ref{sec:Methodology}. We then consider an example system and present proof of concept Monte Carlo results of a fiducial system in Section \ref{sec:dynamics}. We show that this mechanism can potentially result in a much higher EMRI rate in Section \ref{sec:EMRIRate}, and we offer our discussion in Section \ref{sec:Diss}.

\section{Methodology and system set up}\label{sec:Methodology}
\begin{figure}
  \begin{center} 
    \includegraphics[width=\linewidth]{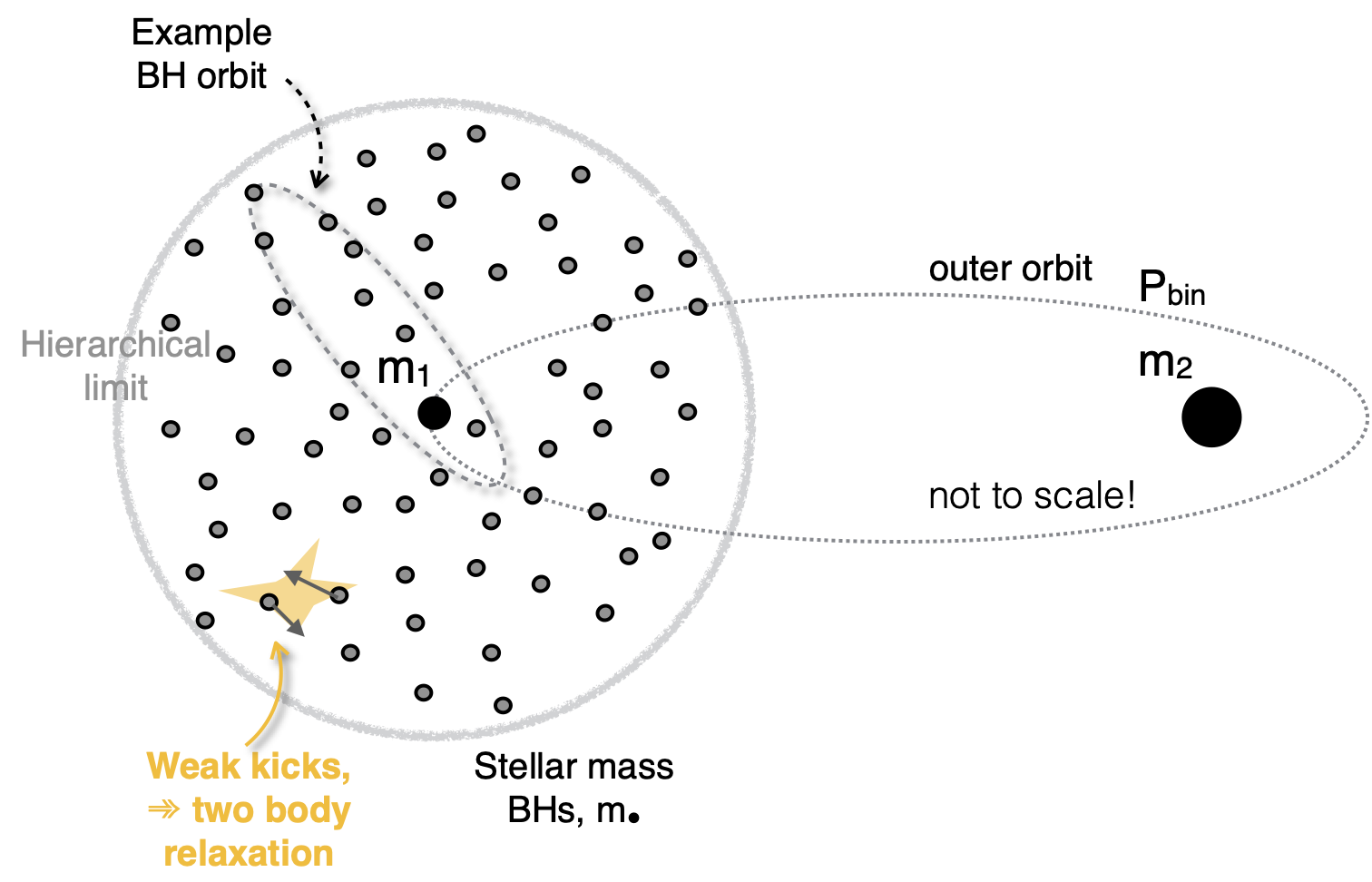} \\
    \includegraphics[width=\linewidth]{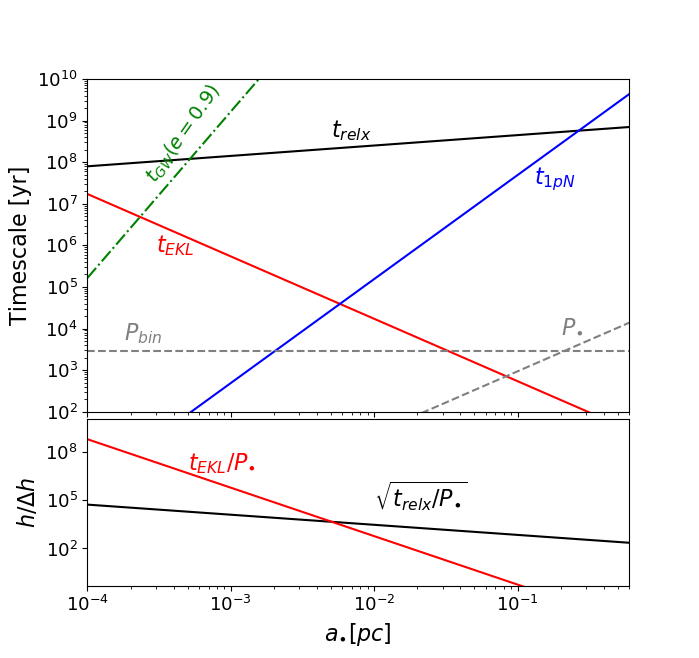}
  \end{center} 
  \caption{  \upshape {\bf Top panel:} An illustration of the system. {\bf Middle panel:} An example of the timescales in the system.  We consider a SMBH of mass $m_1=10^7$~M$_\odot$ with a population of $10$~M$_\odot$ BHs. The period of the BHs around $m_1$ is shown by the dashed line (labeled $P_{\bullet}$), and the associated 1pN precession is shown as the blue line, labeled $t_{1pN}$, according to Equation (\ref{eq:t1PN}). The weak interactions between the BHs  results in the two-body relaxation timescale (see Eq.~(\ref{eq:trelx})), shown by the black line. We also consider a SMBH companion with  $m_2=10^9$~M$_\odot$, at $1$~pc separation (note that $m_1<m_2$ in this configuration). The period of the SMBH binary is shown as the dashed line labeled $P_{\rm bin}$, and the resulting EKL timescale is the red line labeled $t_{\rm EKL}$, see Equation (\ref{eq:tEKL}. {\bf Bottom panel:} Proposed alternative to quantify the relative importance of the two-body relaxation processes compared to EKL. We consider $h/\Delta h$ (the relative change in angular momentum `$h$'), due to both two-body relaxation and EKL, where $h/\Delta h_{\rm relx}\sim \sqrt{t_{\rm relx}/P_\bullet}$ for two-body processes, and $h/\Delta h_{\rm EKL}\sim {t_{\rm EKL}/P_\bullet}$ for EKL. See text (\S \ref{sec:revisit}) for more details. } \label{fig:Cartoon} 
\end{figure}
\subsection{Fiducial System}
We consider a system of SMBH binary with masses $m_1$ and $m_2$,  and orbital period $P_{\rm bin}$. Surrounding the primary $m_1$ is a sphere of compact objects at distance $r_\bullet$ and masses $m_\bullet$, where for simplicity we assume the same masses\footnote{Different population may result in slightly different density profiles, see \citet{Aharon+16}.}. Note that here $m_1<m_2$. We emphasize that the physical processes described below are scalable beyond the fiducial system adopted here. In particular,  we expect that two-body relaxation will play a critical role in the EKL process of a population of stars and a wide range of compact object masses surrounding SMBHs. 
The stellar mass BHs ($m_\bullet$) density profile $\rho(r_\bullet)$ is calibrated by the $M-\sigma$ relation \citep{Tremaine+02}:
\begin{equation}\label{eq:rho}
    \rho(r_{\bullet}) = \frac{3-\alpha}{2\pi} \frac{m_1}{r_{\bullet}^3}\left(\frac{G\sqrt{m_1 M_0}}{\sigma_0^2 r_{\bullet}}\right)^{-3+\alpha} \ ,
\end{equation}
where $M_0=10^8$~M$_\odot$, and $\sigma_0=200$~km~sec$^{-1}$, {are scaling factors. Below we adopt a \citet{Bahcall+76} profile, i.e., $\alpha=1.75$.}  Note that these values have been slightly modified recently \citep[e.g.,][]{van_den_Bosch16,McConnell+13}.  However, it does not affect the underlying physical processes described below, and only may slightly change the relaxation timescale (see Section~\ref{sec:twobody}).   

Each BH ($m_\bullet$) undergoes eccentricity and inclination excitations due to the far away SMBH companion ($m_2$) according to the EKL mechanism. Additionally, general relativity effects induce precession and can also circularize and shrink the orbit through gravitational wave (GW) emission. Finally, collective relaxation interactions with the sea of objects in the sphere of influence tend to change the angular momentum and energy of the orbit by an order of themselves over long timescales. Below we specify these different physical processes and outline the methodology of including them in our analysis. 

\subsection{Three body secular analysis}
We solve the hierarchical three body secular equations up to the octupole-level of approximation \citep[see for complete set of equations][]{Naoz+11sec}.  The timescale of the lowest order of approximation, namely the quadrupole, is \citep[e.g.,][]{Antognini15} estimated by
\begin{equation}\label{eq:tEKL}
    t_{\rm EKL} \sim \frac{16}{30\pi} \frac{m_1+m_\bullet + m_2}{m_2}\frac{P_{\rm bin}^2}{P_\bullet}(1-e_{\rm bin}^2)^{3/2} \ , 
\end{equation}
where $P_{\rm bin}$ and $e_{\rm bin}$ are the period and eccentricity of the SMBH binary respectively, and $P_\bullet$ is the period of the stellar mass black hole around $m_1$. We show this timescale in Figure \ref{fig:Cartoon}.

\subsection{General relativity and Gravitational Waves}
The 1st post Newtonian effects induced by $m_1$ cause $m_\bullet$ to precess on a characteristic timescale
\begin{equation}\label{eq:t1PN}
  t_{\rm 1pN} \sim \frac{P_\bullet c^2 a_\bullet(1-e_\bullet^2)}{6\pi G (m_1+m_\bullet)}
\end{equation}
where $c$ is the speed of light. When this timescale is shorter than the quadrupole timescale from Equation (\ref{eq:tEKL}) eccentricity excitations are typically suppressed \citep[e.g.,][]{Ford00Pls,Naoz+12GR,Will+17,Lim+20}. However, when these two timescales are similar, the precession may excite eccentricities and even re-trigger the EKL behaviour of extreme eccentricity and inclination flips \citep{Naoz+12GR}, by destabilizing the quadrupole level resonance \citep[see,][]{Hansen+20}. The timescales of $m_\bullet$ for a fiducial system are shown in Figure \ref{fig:Cartoon}. 

We include in our calculations both 1st post Newtonian effects from the primary $m_1$ and the secondary $m_2$. As mentioned in \citet{NaozSilk14} and \citet{Li+15}, we choose to focus on the BHs around the less massive SMBH to minimize the part of the parameter space in which 1st pN precession suppresses the EKL's eccentricity excitations. As we highlight below, in the presence of two-body relaxation this suppression is minimized. 

In addition to 1pN precession we also include the shrinking and circularization of the stellar BH orbit due to gravitational wave emission following \citet{Peters+63}. The characteristic timescale to merge an EMRI is:
\begin{eqnarray}
    t_{\rm GW} &\sim&  5.8\times 10^9 ~{\rm yr}\left(\frac{m_1}{10^6~{\rm M}_\odot} \right)^{-2} \left(\frac{m_\bullet}{10~{\rm M}_\odot} \right)^{-1} \left(\frac{a_\bullet}{10^{-4}~{\rm pc}} \right)^4 \nonumber \\ 
    &\times& f(e_\bullet)(1-e_\bullet^2)^{7/2} \ ,
\end{eqnarray}
where $f(e_\bullet)$ is a function of $e_\bullet$ and for all values of $e_\bullet$ is between $0.979$ and $1.81$, \citep[][]{Bla+02}. We show this timescale for our fiducial system in Figure \ref{fig:Cartoon} for $e_\bullet=0.9$.

\subsection{Two-body relaxation}\label{sec:twobody} 
Scattering relaxation interactions of a target black hole with the sea of objects are modeled by considering the  two-body relaxation timescale \citep[e.g.,][]{Binney+TremaineBook}: 
\begin{equation}\label{eq:trelx}
    t_{\rm relx} = 0.34 \frac{\sigma^3}{G^2 \rho\langle m_{\rm scat} \rangle\ln\Lambda} \ ,
\end{equation}
where $\langle m_{\rm scat}\rangle$ is the mass of the average scatterer, $\sigma$ is the velocity dispersion of BHs around the SMBH
\begin{equation}
    \sigma^2=\frac{Gm_1}{r_\bullet(1+\alpha)} \ ,
\end{equation}
where $\alpha$ is the slope of the density profile. The coulomb logarithm is:
\begin{equation}
    \Lambda = \frac{r_\bullet \sigma^2}{2G \langle m_{\rm scat}\rangle}  \ .
    \end{equation}
For simplicity we adopt $\langle m_{\rm scat} \rangle \approx m_\bullet$. However, if $\langle m_{\rm scat}\rangle << m_\bullet$, mass segregation may migrate the BHs inwards.  

The relaxation time from Equation (\ref{eq:trelx}),  is the timescale for a change of energy of the stellar mass BH around the SMBH $m_1$ by an order of its orbital energy, or a change in angular momentum by an order of its circular angular momentum. We show the relaxation timescale in Figure \ref{fig:Cartoon} (solid black line on top), which for large part of the parameter space is much larger than the EKL timescale. As mentioned, this motivated many studies to ignore the contribution of two-body relaxation when considering EKL effects.  

The typical change in the  BH's velocity $v_\bullet=\sqrt{Gm_1 ( 2/r_\bullet - 1/a_\bullet)}$ due to one encounter is:
\begin{equation}\label{eq:Deltav}
    \Delta v = v_\bullet\sqrt{ \frac{P_\bullet}{t_{\rm relx}}} \ . 
\end{equation}
We model this change as a random walk, applying a single isotropically oriented kick to the BH velocity once per orbit around the SMBH.  Each directional component of this 3D kick is drawn from a Gaussian distribution with a zero average and a standard deviation of $\Delta v/\sqrt{3}$ \citep[see][for a similar approach for binaries around a single SMBH]{Bradnick+17}. 
We assume that the kick is instantaneous at some random phase of the BH's orbit
\begin{equation}
    r_\bullet=\frac{a_\bullet(1-e_\bullet^2)}{1+e_\bullet \cos f_\bullet} \ ,
\end{equation}
where $f_\bullet$ is the true anomaly\footnote{Note that we choose the Eccentric anomaly from a uniform distribution and finding the true anomaly from there.}. Thus, the vector $\vec{r}_\bullet$ in the invariable plane\footnote{Note that the system evolves due to EKL and thus we need to project the separation vector on the invariable plane. For similar analysis see \cite[e.g.,][]{Lu+19}.} can be considered constant during the encounter. See appendix \ref{sec:app} for full set of the two-body relaxation equations. 

\begin{figure}
  \begin{center} 
    \includegraphics[width=\linewidth]{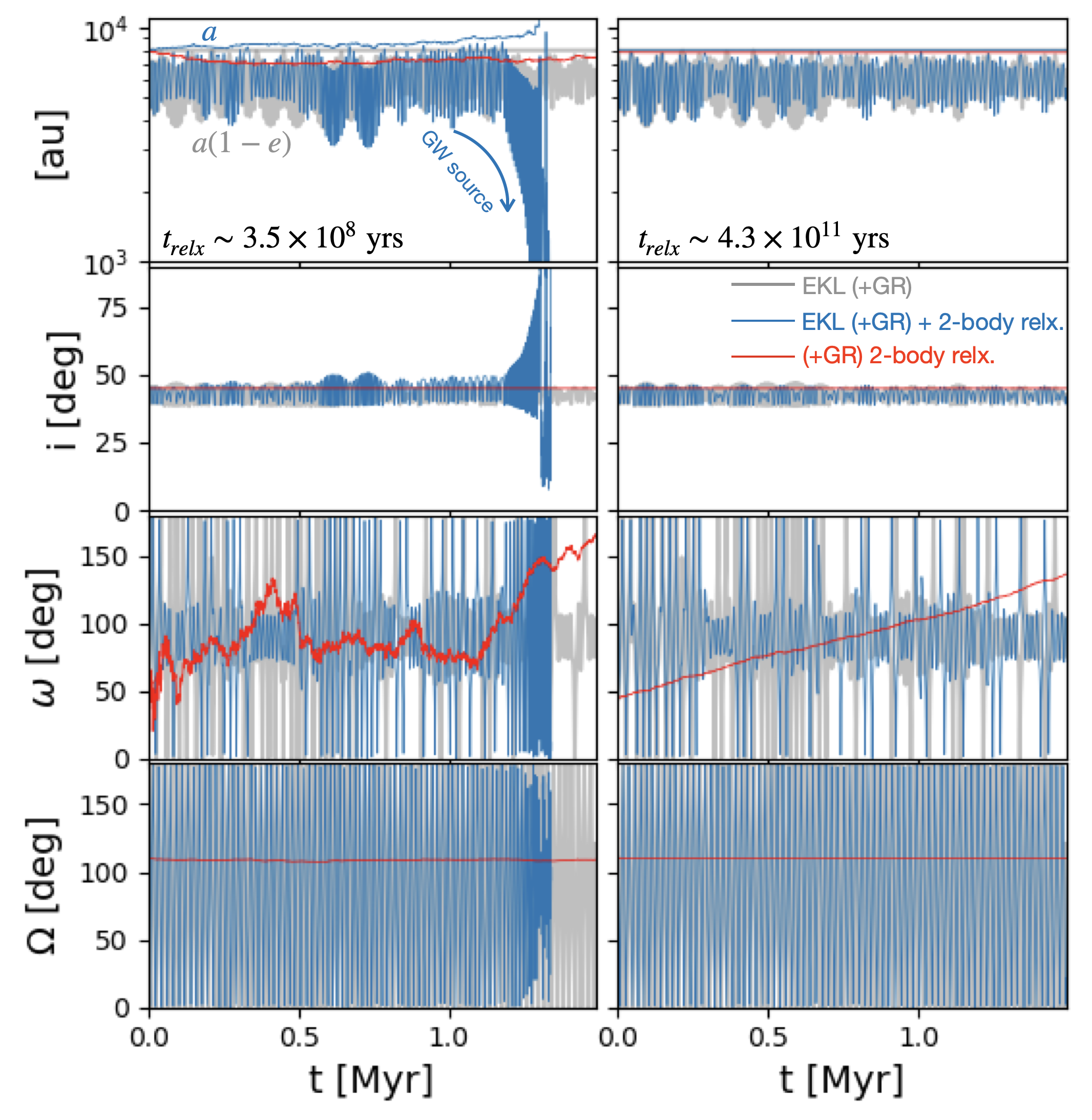}
  \end{center} 
  \caption{  \upshape {\bf Time Evolution of an example system in the presence of different physical processes.} We show, from top to bottom, a stellar mass black hole separation around an SMBH, inclination with respect to the outer perturber, argument of perihapsis, and longitude of ascending nodes.    {\it Left side:} We consider a stellar mass black hole ($m_{\bullet}=10$~M$_\odot$) orbiting an SMBH ($m_1=10^7$~M$_\odot$), at $a_\bullet=8000$~au, initially with $e_\bullet=0.02$, $\omega_\bullet=45^\circ$, $\Omega_\bullet=110^\circ$. We also consider a population of stellar mass black holes around $m_1$, following a \citet{Bahcall+76} profile (i.e., $\alpha=1.75$). We normalize the density profile according to the $m-\sigma$ relation, (see Equation (\ref{eq:rho})), which result in two-body relaxation timescale of $t_{\rm relx}\sim 3.5\times 10^8$~yrs. We show the resulting orbital evolution of the stellar black hole in the thick red line. We also introduce a binary SMBH with mass $m_2=10^9$~M$_\odot$ set on $1$~pc separation, with eccentricity of $e_{\rm bin}=0.7$. The evolution that includes both the two-body relaxation and the EKL from the outer orbit (as well as GR precession on the inner orbit) is shown in thin blue line. {\it As depicted this system reached extreme eccentricities induced by a combination of two-body relaxation and EKL and pushed toward the SMBH, producing a GW source.} We also consider the case of which we ignore the contribution of two-body relaxation processes and consider only the EKL (+GR) in light grey. This system never reached high eccentricity to become an EMRI. 
  {\it At the right side} we consider the same system, only this time we arbitrary increased the relaxation timescale to $4.3\times 10^{11}$~yrs (by assuming scatter masses of $5\times 10^{-3}$~M$_\odot$). As depicted this system qualitatively follows the EKL (+GR) behaviour.     } \label{fig:TimeEvolution} 
\end{figure}

\section{Dynamical evolution}\label{sec:dynamics}
\subsection{Example system and revisiting the time-scale argument}\label{sec:revisit}
The EKL mechanism tends to excite high eccentricities and inclination. However, only about $30\%$ of the parameter space in the aforementioned configuration is available to reach the extreme eccentricities needed to {drive} an object into the black hole, and cross its Schwarzschild radius \citep[e.g.,][]{Li+15,NaozSilk14,Naoz+19}. As an example, we consider in Figure \ref{fig:TimeEvolution} a system whose EKL eccentricity excitations do not result in values sufficient to cross the SMBH's Schwarzschild radius (grey lines in both columns). For this system, the EKL timescale ($t_{\rm EKL}\sim 1.4\times 10^4$~yr) is shorter than the GR precession timescale ($t_{\rm 1pN}\sim 6\times 10^6$~yr). 

However, as can be seen in Figure \ref{fig:TimeEvolution}, left column, a two-body relaxation process combined with EKL results in aggravated EKL eccentricity and inclination excitations. We note that we include GR precession for the inner and outer orbit. The former suppresses the EKL eccentricity excitations when two-body relaxation is not included (gray lines). However, in this example we do not include GW emission. To avoid clutter, GW is included in the Monte Carlo analysis below. In this example (Figure \ref{fig:TimeEvolution}, left column) we consider a black hole population with a \citet{Bahcall+76} distribution (i.e., $\alpha=7/4$). The two-body relaxation timescale from Equation (\ref{eq:trelx}) is $t_{\rm relx}\approx 3.5\times 10^8$~yrs, well above the the EKL timescale (see also Figure \ref{fig:Cartoon}, for this case it's about four orders of magnitude larger). By definition, over the $\approx 1.5$ Myrs run, the relaxation timescale is insufficient to change the angular momentum by an order of itself (because the timescale is shorter than $t_{\rm relx}$ in this case).
However, the combined effect of two-body relaxation and EKL results in higher eccentricity and inclination amplitude modulations. 

In fact, the eccentricity excitations were large enough to {drive} this stellar mass BH onto the SMBH, thus forming an EMRI. The higher eccentricity values reached are correlated with the BH semi-major-axis slightly drifting to higher values, due to two-body relaxation, thus getting closer to the secondary SMBH ($m_2$). This process yields a shorter EKL timescale (recall Eq.~(\ref{eq:tEKL}) dependency on the inner orbital period). Furthermore, as the inner orbit gets closer to the secondary SMBH, the octupole-level of approximation dominates more. This behaviour is expressed by the pre-factor of the  octupole-level Hamiltonian $\epsilon$ \citep[e.g.,][]{LN11}:
 \begin{equation}\label{eq:epsilon}
     \epsilon = \frac{a_\bullet}{a_{\rm bin}}\frac{e_{\rm bin}}{1-e_{\rm bin}^2} \ .
 \end{equation}
 Thus, as $a_\bullet$ increases, so does $\epsilon$, which excites the eccentricity of the BH toward larger values  \citep[e.g.,][]{Li+13,Li+14}.

The obvious questions from this result are why these diffusion processes create such a large effect, and will it always happen regardless the value of $t_{\rm relx}$. The answers to both of these questions can be understood by examining Equation (\ref{eq:Deltav}), which suggests that $h/\Delta h|_{\rm relx} \sim \sqrt{t_{\rm relx}}$, where $h$ is the angular momentum and $\delta h$ is the change of the angular momentum due to a small kick over the particle orbit around $m_1$. However, the angular momentum changes due to the EKL are $h/\Delta h|_{\rm EKL} \sim {t_{\rm EKL}}$ \citep[e.g.,][]{Naoz+11sec}. Thus, effectively,  we should compare $\sqrt{t_{\rm relx}/P_\bullet}$ to $t_{\rm EKL}/P_\bullet$. We show this comparison in Figure \ref{fig:Cartoon}, bottom panel, where we compare $h/\Delta h$, due to the different processes. Using this picture, it is clearer that two-body relaxation is relevant to a large part of the parameter space. 

In the example depicted in the left column of Figure  \ref{fig:TimeEvolution}, even though $h/\Delta h|_{\rm relx} > h/\Delta h|_{\rm EKL}$,  it is only by a factor of $20$, which yields this cumulative effect (examining the bottom panel in Figure \ref{fig:Cartoon}, helps clarify the comparison between the two effects). About an order of magnitude difference can still lead to a significant cumulative effect. This behavior is similar to the way that GR precession destabilizes the quadrupole resonance, even when it's timescale is much longer than the quadrupole level \citep[e.g.,][]{Naoz+17,Hansen+20}.   We note of course that for this system, two-body relaxation effects would have eventually change the energy and angular momentum of the orbit by an order of themselves, regardless of EKL. However, this does not guaranteed an orbit that will plunge onto $m_1$. In our case we have adopted a \citet{Bahcall+76}, i.e., $\alpha=7/4$, which results in zero net flux, thus, the BHs are expected to undergo diffusion, but not preferentially migrate.  

We emphasize that the two-body relaxation effect {on the orbital configuration} is indeed small compared to the long-term EKL eccentricity excitation. This is highlighted in Figure \ref{fig:TimeEvolution} for the two-body relaxation-only case (red lines), which does not excite the eccentricity to any meaningfully high values during the simulation run-time. Instead, the BH simply undergoes diffusion in its energy and angular momentum. However, since the EKL is sensitive to the orbital configuration, the diffusion in energy and angular momentum due to two-body relaxation can still contribute to large effects on the BH orbit. If the small changes in the orbit's energy and angular momentum can cause a change of the angular momentum of about $10-15\%$, the effects on EKL are substantial.  

For comparison, we consider the same system in Figure  \ref{fig:TimeEvolution} (right column), only this time we artificially increased the relaxation timescale, for illustration purposes.  In this example $t_{\rm relx}\approx 4.3\times 10^{11}$~yrs, which is also longer than the lifetime of the system, and the BH simply undergoes diffusion. As clearly depicted in the Figure, the diffusion in this system is insignificant and does not trigger larger EKL effects.  Furthermore, in this example, we find that $h/\Delta h|_{\rm relx} \approx 700 \times h/\Delta h|_{\rm EKL}$. Thus, the relaxation effects, according to this comparison, results in a negligible change. In this panel, we again over-plot the two-body relaxation-only effect (+$1$pN), as shown by the thick red lines. Note that the apparent drift in $\omega$ in this case is due to the $1$pN precession, a similar drift, is depicted in the left column, only modulated by the diffusion processes.
 
As depicted in the bottom two panels in Figure \ref{fig:MonteCarlo}, the nominal suppression of eccentricity excitations due to $1$pN precession does not take place. To guide the eye we have outline the $t_{\rm EKL}=t_{1pN}$ line for a $e_\bullet=2/3$.  Indeed, without two-body relaxation processes, eccentricity excitations are suppressed in the presence of GR precession \citep[e.g.,][]{Ford00Pls,Naoz+12GR}. However, the small kicks result in a diffusion, thus allowing the eccentricity excitation to take place over a wide range of the parameter space.    
 
Lastly, a striking feature of Figure \ref{fig:TimeEvolution} is that in the presence of two-body relaxation the system moves in and out libration regime, not in-sync with EKL. The resonant angle, $\omega$, is known to change from libration to circulation in EKL \citep[e.g.,][]{Li+14,Hansen+20}. However, as depicted, the diffusion process changes these processes, even when the two-body relaxation effects are insignificant. These small kicks allow the (already chaotic) system to transfer zones. 
 
\subsection{Monte Carlo Proof-of-concept} 

 As mentioned, two-body relaxation processes are often neglected when analyzing the EKL like systems. On the other hand, EKL is often neglected when considering objects {sinking} onto a SMBH. Here, we qualitatively show  the importance of combining these two processes. We consider the system highlighted in Figure \ref{fig:Cartoon}, of $m_1=10^7$~M$_\odot$ and $m_2=10^9$~M$_\odot$ with a binary separation of $a_{\rm bin}=1$~pc and eccentricity of $e_{\rm bin}=0.7$. We populate the area of $m_1$ with $1000$ stellar mass BHs, adopting a \citet{Bahcall+76}, i.e., $\alpha=7/4$, profile.  We also adopt a thermal distribution for the stellar mass black holes, and a mutual inclination that is taken from isotropic distribution (i.e., uniform in $\cos i$). The argument of perihapsis and longtitue of ascending nodes are taken from a uniform distribution between $0-2\pi$. 
 
 \begin{figure}
  \begin{center} 
    \includegraphics[width=\linewidth]{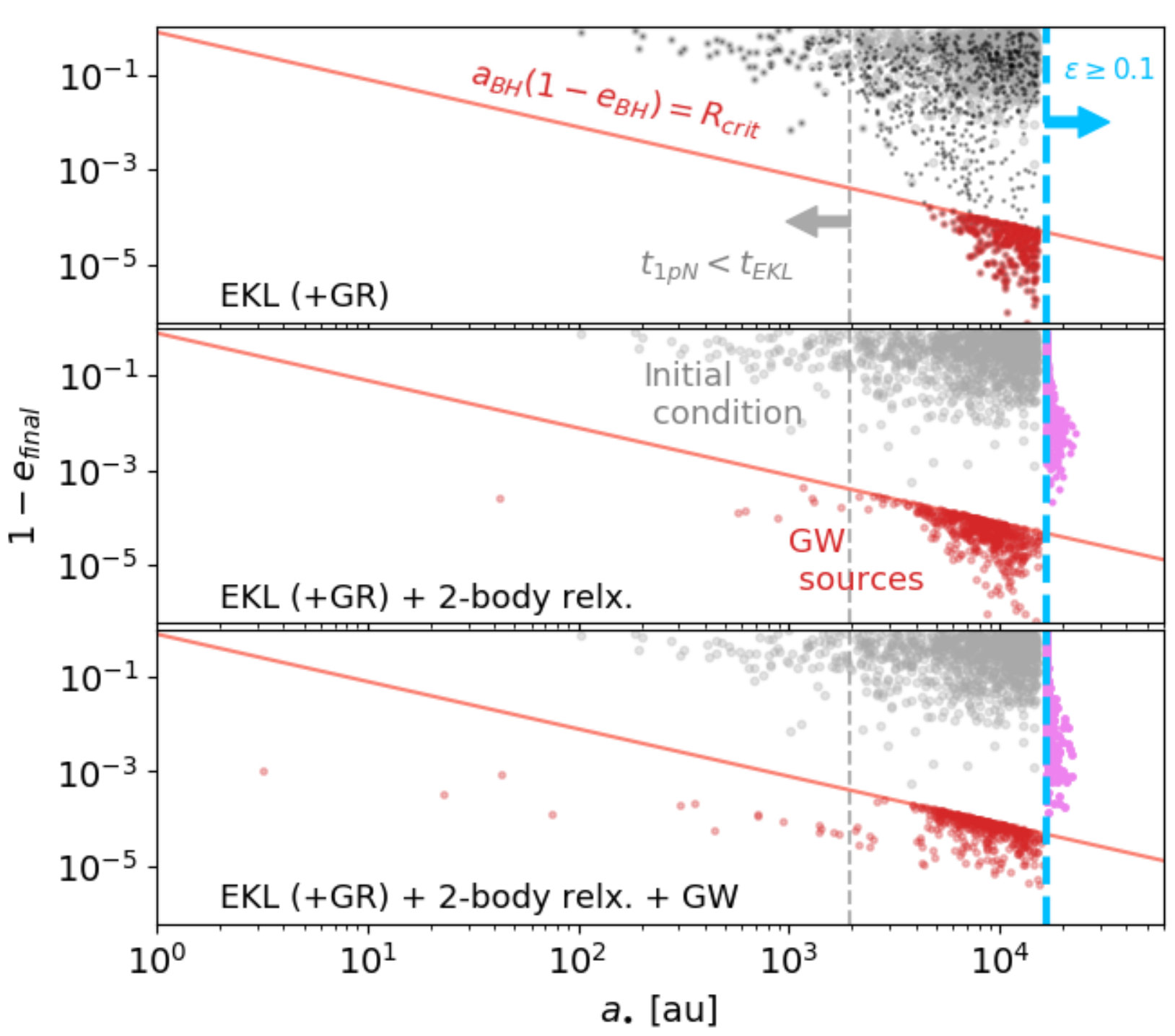}
  \end{center} 
  \caption{  \upshape {\bf Monte-Carlo results.} As a proof of concept, we consider a system composed out of  $m_1=10^7$~M$_\odot$ and $m_2=10^9$~M$_\odot$ with a binary separation of $a_{\rm bin}=1$~pc and eccentricity of $e_{\rm bin}=0.7$. We present three runs, of $1000$ particles each. We consider the following processes: (top) EKL + GR, (middle) EKL + GR + two-body relaxation, (bottom)   EKL + GR + two-body relaxation + GW. The initial conditions are the same at each run and are shown in grey in each panel. Red line marks the limit of crossing $R_{\rm crit}$, thus a system that ended up below the line is marked as a potential GW source, i.e., EMRI.    } \label{fig:MonteCarlo} 
\end{figure}
 
 In Figure \ref{fig:MonteCarlo} we present the results of  $3$ set of simulations of $10^3$ particles each, while adopting the following physical processes (top) EKL + GR, (middle) EKL + GR + two-body relaxation, (bottom)   EKL + GR + two-body relaxation + GW. The light grey point in each panel represent the initial conditions (which are identical in each panel). We have three stopping conditions:
 \begin{enumerate}
     \item The simulation reaches $10^9$~yrs (depicted as black small points). This result only happen in the  EKL + GR run (top panel), where about $69\%$ of the systems have survived throughout the EKL + GR simulation \citep[this is consistent with the results for dark matter particles, by][]{NaozSilk14}. 
     \item  The stellar mass BH pericenter crossed a critical distance,  which we adopt as $R_{\rm sch}=8 Gm_1/c^2$, \citep[following][]{NaozSilk14,Naoz+19}, which is inside the {inside the Kerr black hole's inner-most retrograde stable orbit.} 
     These are represented by red points below the solid line. We label them as ``GW sources.''  In the EKL + GR, about $31\%$ of all systems crossed the the critical radius, while $50\%$ ($53\%$) of all systems in the EKL + GR + two-body relaxation (+GW) run have ended up as GW sources. 
     \item  The BH semi-major axis changed due to two-body relaxation such that $\epsilon>0.1$ (pink points, to the right of the dashed line). This is only possible when the two-body relaxation is turned on. 
 \end{enumerate}

While it is clear that the systems whose pericenter crossed $R_{\rm sc}$, are GW sources (i.e., EMRI candidates), it may be less obvious to understand what is the outcome of those with $\epsilon >0.1$. We emphasize that this condition for hierarchy is based on the octupole pre-factor and therefore is somewhat arbitrary \citep[e.g.][]{LN11}. Furthermore, it was suggested in \citet{Bhaskar+21} that violating this role often results in even higher eccentricities. Thus, we refer to those systems as possible EMRIs candidate as well\footnote{{Note that systems that crossed the Roche limit (or the Hill Sphere) of the secondary may also be considered as systems that  descend toward the SMBH (either the primary or secondary) following \citet{Chen+08,Chen+09,Chen+11}. Similar arguments were done for systems for which $\epsilon >0.1$ \citep[e.g.,][]{Bhaskar+21}. Furthermore, Zhang et al. in prep. showed that even in the case of this Roche limit crossing of a tertiary the system may not change its  energy or angular momentum at the order of itself for long timescales. In other words, the system may still considered ``stable'' and the eccentricity may continue to increase via EKL.  Thus, the combined effect of EKL + two-body relaxation processes may continue to occur for BHs for which $\epsilon>0.1$, until resulting in possible EMRIs (see Appendix \ref{sec:SNR}). }  }.  We find that between $\approx 50-100\%$ of the BHs (corresponding to pericenter smaller than $R_{\rm crit}$, to $\epsilon>0.1$) become a GW source.   

{For peri-centers smaller than $R_{\rm crit}$,  Kerr geometry may cause the BHs to spend a lot of time on the SMBH's ergosphere \citep{Schnittman15} where GW emission may shrink their separations. Furthermore, special relativity effects should also be taken into account \citep{Yunes+08,Berry+13}.    }

As can be seen from Figure \ref{fig:MonteCarlo}, the combination of EKL with two-body relaxation allows the system to access a larger part of the parameter space, thus triggering the EKL mechanism. In general, the number of objects that undergo high eccentricity excitation depend on the density distribution \citep[e.g.,][Mockler et al.~in prep.~]{Li+15}. 
Moreover, since the two-body relaxation timescale is highly sensitive to the density profile \citep[i.e., $\alpha$, see for example, Figure 1 in][]{Rose+20} we expect that the efficiency of the combined system will depend on the underlying density distribution (see   Melchor et al. in prep.).

\section{EMRI Rate estimation}\label{sec:EMRIRate}

\begin{figure}
  \begin{center} 
    \includegraphics[width=\linewidth]{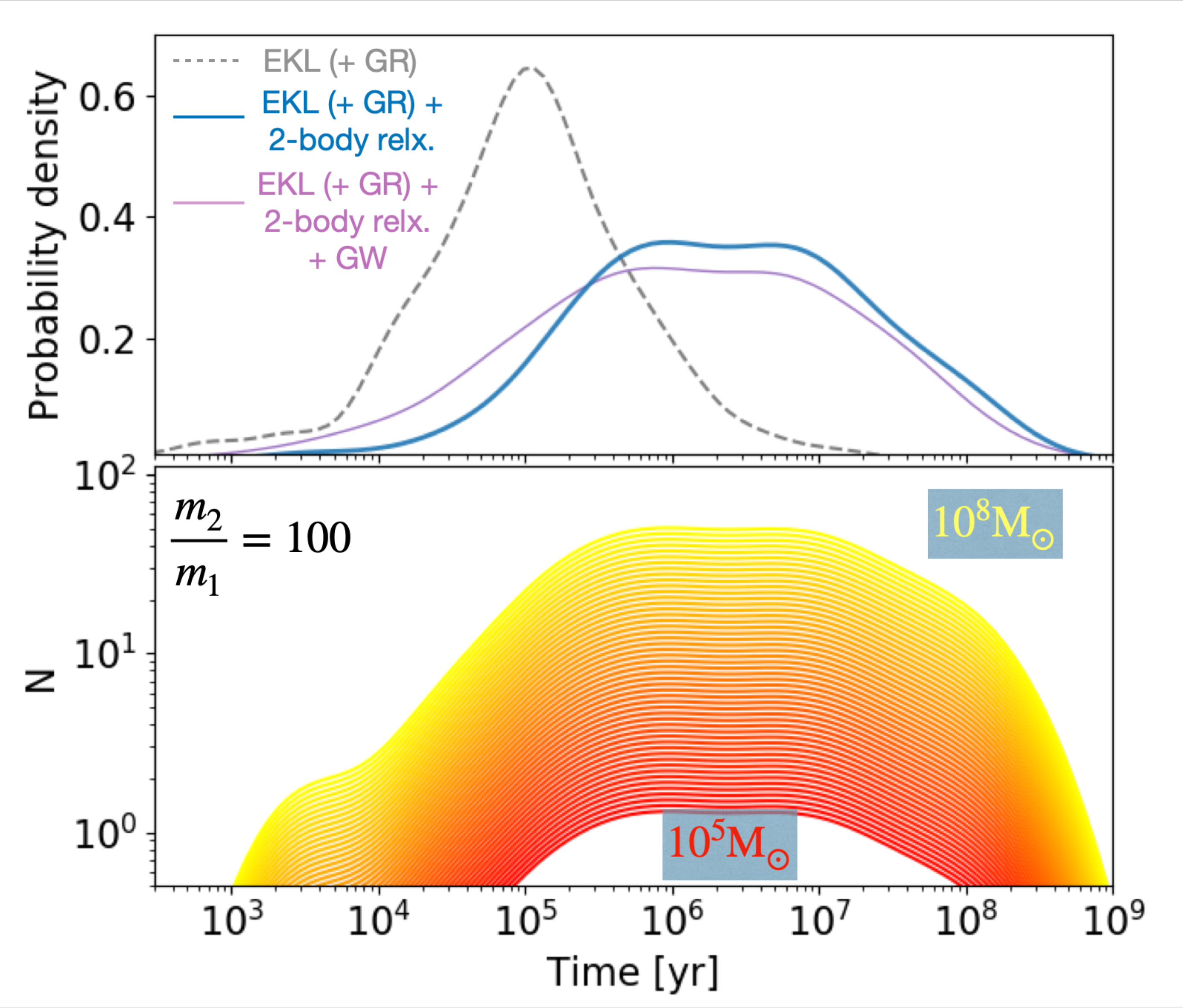}
  \end{center} 
  \caption{  \upshape {\bf Scaling relation proof of concept.}  {{\it Top panel},  shows the probability density of the BHs that cross $R_{\rm crit}$ and thus becomes a GW source, as a function of time,} in each of the three simulations from Figure \ref{fig:MonteCarlo}.  Note that the statistical difference between the two probability densities that include the two-body relaxation is negligible. In particular, the two-sample Kolmogorov-Smirnov test  does not rejects the null hypothesis, that both has the dame distribution, at the $20\%$ significance level. {\it Bottom panel} shows the number of stellar mass BHs that became GW sources as a function of time, for a range of primary masses between $10^5-10^8$~M$_\odot$ (from bottom to top). In generating this estimate we have assumed constant $t_{\rm EKL}$,  constant mass ratio, and that the maximum distance for stellar mass BHs corresponds to $\epsilon=0.1$, see Eq.~(\ref{eq:epsilon}). We estimate the number of stellar mass BHs using the $M-\sigma$, relation (see text for more details). 
    } \label{fig:Acc} 
\end{figure}

The rate estimation is very sensitive to the steady state number of BHs around the SMBH. It varies over three orders of magnitude between the various assumptions for EMRIs formation processes \citep[e.g.,][]{Freitag01,Hopman+05,Hopman09,Amaro+11,Aharon+16,Bar-Or+16,Babak+17}. Thus, here we aim to highlight the efficiency of the proposed mechanism by utilizing the $M-\sigma$ relation for the number of BHs. We then compare to similar approaches in the literature for the two-body relaxation process. 

The EKL-only runs compared to the ones with two-body relaxation processes yield a significantly different flux of GW source formation. This is shown in the top panel of Figure \ref{fig:Acc}, where a striking feature is the EKL (+GR)-only result. This feature is consistent with a {\it``burst''-like} behavior that depletes the stellar mass BHs, which could otherwise become GW sources \citep[a similar behavior was found for TDEs and dark matter participle depletion][]{Li+15,NaozSilk14}. Thus, for a relatively short time ($6\times 10^5$~yr, corresponding to the width of the distribution), the rate is high, but on the timescale it takes to replenish the stellar mass BH population, the rate is low.  
Replenishment of BHs can take place via mass segregation, which brings BHs in from the sphere of influence \citep[e.g.,][]{Hopman+06seg}. The corresponding timescale at the order of the two-body relaxation timescale up to a factor of the mass ratio between the BHs and the stars.  Another source of replenishment is star formation, which for our galactic center is estimated to occur every few$\times 10^6$~yr \citep{Lu+13}. Unlike the EKL (+GR)-only result, the inclusion of two-body relaxation expands the timescales at which GW sources can form, thus allowing for the replenishment of stellar-mass BHs to take place\footnote{{Note that in these cases, during the long timescales the SMBH binary's separation is expected to shrink, yielding an enhancement to the EMRI rate \citep[e.g.,][]{Iwasa+16}. The inclusion of this effect is beyond the scope of this paper.  }}.  

To estimate the number of black holes, $n_{\rm BH}(\leq r_{\bullet})$, within a distance $r_{\rm max}$ we use the $M-\sigma$ relation:
\begin{eqnarray}\label{eq:nBH}
    n_{\rm BH}(\leq r_{\rm max}) &=&  f_{\rm BH}\frac{ M(\leq r_{\rm max})}{\langle m_\star\rangle}  \\ &=&2 f_{\rm BH} \frac{ m_1}{\langle m_\star \rangle}\left(\frac{G\sqrt{m_1 M_0}}{\sigma_0^2 r_{\bullet}}\right)^{-3+\alpha} \nonumber \ ,
\end{eqnarray}
where $M(\leq r_{\rm max}) = \int_0^{r_{\rm max}}\rho (r') 4\pi r'^2dr'$ and $\rho$ is the density profile form Equation (\ref{eq:rho}). Furthermore, $\langle m_\star \rangle$ is the average mass of the stars and $f_{\rm BH}$ is the fraction of BHs from the overall stellar population, where we adopt $f_{\rm BH} = 3.2\times 10^{-3}$ \citep[e.g.,][]{Aharon+16}. In our fiducial system $r_{\rm max}=0.07$~pc, which corresponds to the $\epsilon=0.1$, and the number of BHs within this radius is about $330$.

As highlighted in previous studies, it is straightforward to scale the system to a wide range of primary masses, for a constant mass ratio, while holding the quadrupole timescale (Eq.~\ref{eq:tEKL}) constant\footnote{Note that we limit our analysis to systems for which $t_{\rm EKL}<t_{\rm relx}$, to allow for a the behavior outlined in Figure \ref{fig:MonteCarlo} to take place.} and considering the number of BHs up to $r_{\rm max}$, for $\epsilon=0.1$, \citep[e.g.,][]{NaozSilk14,Naoz+19}.   Thus, in Figure  \ref{fig:Acc}, bottom panel, we show the number of stellar mass BHs that are {sunk} onto the SMBH, for the run that includes all of the aforementioned physical processes.  In this scaling, proof of concept, $r_{\rm max}$ is then mass dependent and it takes the following form:
\begin{eqnarray}\label{eq:rmax}
r_{\rm max}& = &\left(\frac{15}{16}\right)^{2/3} t_{\rm EKL}^{2/3} q^{-2/3} \left(\frac{\epsilon}{e_{\rm bin}}\right)^2(1-e_{\rm bin}^2)(Gm_1)^{1/3} \nonumber \\
&\approx& 0.07~{\rm pc}\left(\frac{t_{\rm EKL}}{930~{\rm yr}} \right)^{2/3} \left(\frac{q}{0.01}\right)^{-2/3}\left(\frac{\epsilon}{0.1}\right)^2 \nonumber \\ 
&\times& \left(\frac{e_{\rm bin}}{0.7}\right)^2\left(1-\bigg[\frac{e_{\rm bin}}{0.7} \bigg]^2 \right)\left(\frac{m_1}{10^7~{\rm M}_\odot} \right) \ , 
\end{eqnarray}
where $q=m_1/m_2$ is the mass ratio. We note that both in Figure \ref{fig:MonteCarlo} and below we refer to these objects as GW sources, and EMRIs. 

The EMRI rate is then estimated by:
\begin{equation}\label{eq:Gamma}
    \Gamma \approx \Gamma_{\rm EKL} \times f_{\rm EKL} \times f_{\rm EMRI} \times n_{\rm BH}(\leq r_{\rm max})  \ ,
\end{equation}
{where $f_{\rm EMRI}$ is the fraction of systems that may become an EMRI rather than a plunged orbit,} $f_{\rm EKL}$ is the fraction of systems that have their eccentricity excited to cross $R_{\rm sch}$ and $\Gamma_{\rm EKL}$ is the rate estimated in the simulation. We estimate the latter by calculating the average accretion time and estimating $\pm 68\%$ of it  from our fiducial simulations (i.e., taking $1\sigma$ of the accretion time, estimated from Figure \ref{fig:Acc}) and normalized to the range of primary masses as described above (see Figure \ref{fig:Acc} bottom panel). As highlighted in Figure \ref{fig:MonteCarlo}, a large fraction  of systems {sink} onto the SMBH when both EKL and two-body relaxation operate, i.e., $f_{\rm EKL}\sim 0.5 - 1$.  

{In  Appendix \ref{sec:SNR}, we estimate the fraction of systems that are likely to appear within the LISA band ($f_{\rm EMRI}$). Roughly speaking one divides between plunging orbits which may be characterized with a short GW burst and EMRIs that have many to a few cycles before merging with the SMBH \citep[e.g.,][for further discussion]{Rubbo+06,Yunes+08,Berry+13}. In the former case, special relativity correction may need to be included \citep[e.g.,][]{Yunes+08}. Additionally, we note that the pN treatment unitized here may break down around a rotating SMBH, because the stellar mass BHs are expected to spend a lot of time close to the SMBH’s ergosphere, before continuing on their original trajectory  \citep{Schnittman15}. Thus, GW emission may alter their orbit. Therefore, the distinction between plunging and cycling orbit represents a larger problem in this field. 

Based on the above distinction (see Appendix \ref{sec:SNR} for more details), we find that about $40\%$ of the systems may be defined as EMRIs. In  Appendix \ref{sec:SNR} we also present possible SNR of an example system. Note that the fraction of systems that may end up in the LISA band may depend on the distance of the source.    }

\begin{figure}
  \begin{center} 
    \includegraphics[width=\linewidth]{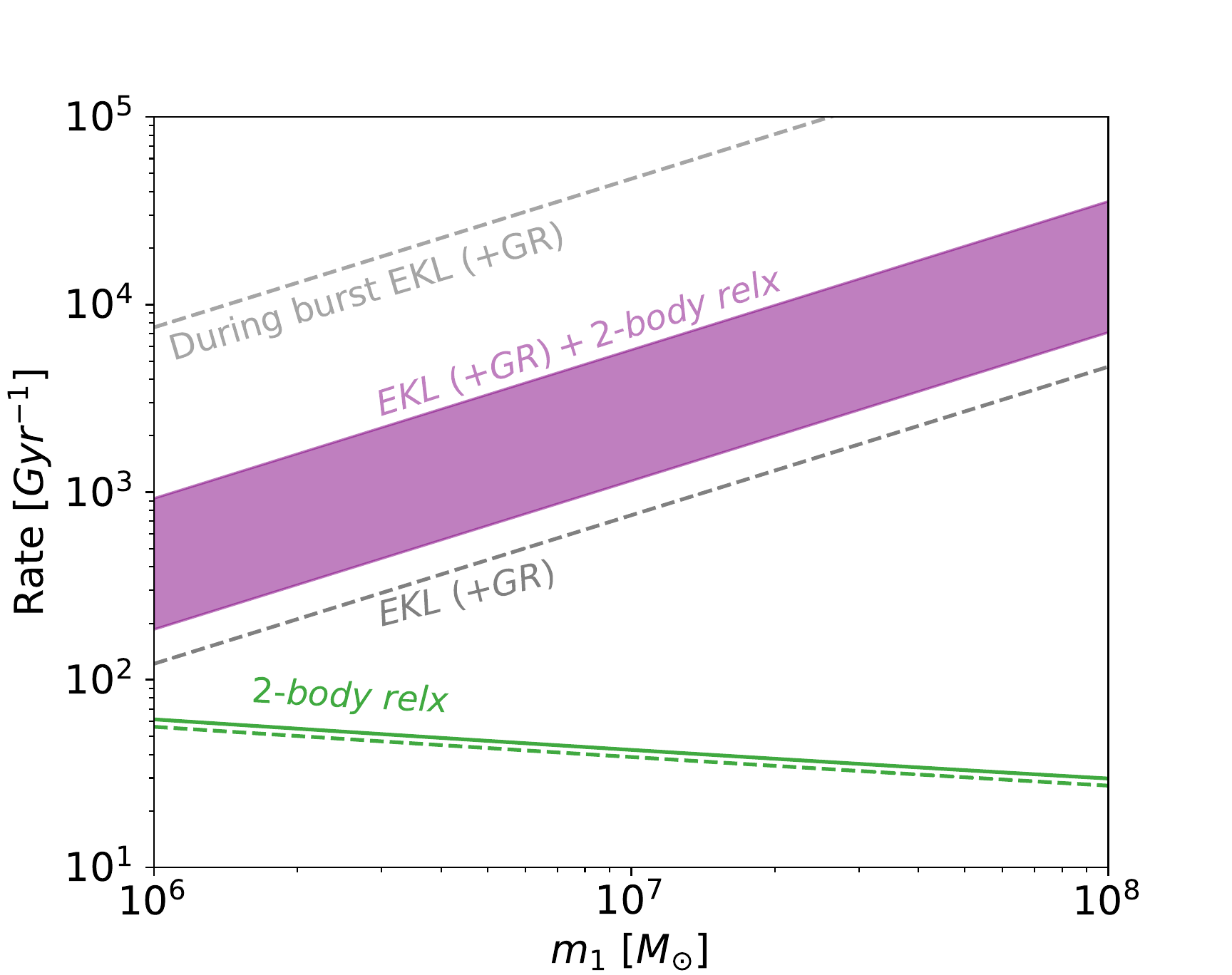}
  \end{center} 
  \caption{  \upshape {\bf A comparison of EMRI formation rate for consistent number of BHs.   } We consider the case which includes EKL (+GR) + two-body relaxation estimated rate from Equation (\ref{eq:Gamma2}). We compare to the EKL (+GR) only runs, where we consider during burst (light dashed line), or over sufficient replenishment time (dark dashed line). The latter is loosely estimated by assuming star formation episode and life time of stars to be about $50$~Myr. Finally, we depict the EMRI rate for the number of BH limited up to $r_{\rm max}$ (sphere of influence), shown as dashed (solid) line.  
   } \label{fig:EMRIRate} 
\end{figure}

Using our scaling relation, and the number of BHs from Eq.~(\ref{eq:nBH}),   the rate is proportional to the mass of the SMBH primary in the following way: \begin{eqnarray}\label{eq:Gamma2}
    \Gamma &\approx&  \Gamma_{\rm EKL} \times f_{\rm EKL} \times  f_{\rm EMRI} \times 2 f_{\rm BH} \times m_1^{(3+\alpha)/6}  \\ &\times & 
    \left(\frac{\left(e_{\rm bin}/\epsilon\right)^2}{1-e_{\rm bin}^2} \frac{G^{2/3}M_0^{1/2}}{\sigma_0^2} \right)^{-3+\alpha}\left(\frac{15\times t_{\rm EKL}}{16 q}\right)^{2(3-\alpha)/3}  \ . \nonumber
\end{eqnarray}
Thus, for the scaling relation chosen in this proof of concept, where $\alpha=7/4$, $q=0.01$, $e_2=0.7$, and  $t_{\rm EKL}$ are constant, the rate is proportional to $m_1^{7/8}$. We show this rate in Figure \ref{fig:EMRIRate}, as the shaded band  for the following limits:  $f_{\rm EKL}\times f_{\rm EMRI} = 1-0.2$, where $f_{\rm EKL}$ corresponds to having $50\%$  ($100\%$)  from the total number of available BHs become EMRIs and $f_{\rm EMRI}=1-0.4$ (see Appendix \ref{sec:SNR}).   

We also depict the EKL (+GR) - only case, during burst (thin dashed line), and the average over replenishment time, taken to be few$\times 10^7$~yr. The EKL (+GR)-only scenario may represent a shallow density distribution ($\alpha\approx  1$) for the BHs, where the two-body relaxation effect is longer and thus can be neglected.   However, the density distribution of BH is expected to be steep \citep[e.g.,][]{Bahcall+76}, and therefore, as highlighted here two-body relaxation processes cannot be ignored. We thus, predict the shaded band as the rate from SMBH binary.

For comparison, we examine the EMRI rate due to only two-body relaxation. For consistent comparison, we only consider the rate due to the ``available'' BHs up to $r_{\rm max}$ [i.e., $n_{\rm BH}(\leq r_{\rm max})$, Eq.~(\ref{eq:nBH})]. The rate is proportional to the number of BHs over the two-body relaxation timescale. However, as highlighted by \citet{Hopman+05}, the onset of GW dissipation does not necessarily correspond with the emission of detectable GW emission.  Thus, following \citet[][equation 31.]{Hopman+05} we write the two-body relaxation EMRI rate as:
\begin{equation}\label{eq:RateRelx}
     \Gamma_{\rm relx}(\leq r_{\rm max}) \approx \frac{n_{\rm BH}}{t_{\rm relx}(r_{\rm max})\ln{\delta J}} \left(\frac{a_c}{r_{\rm max}} \right)^{9/2-2\alpha} \ ,
\end{equation}
where $\delta J$ is the ratio of the maximal circular angular momentum at $a_c$, compared to the angular momentum at the loss cone. Finally, $a_c$ is the critical semimajor axis at which the angular momentum relaxation time is equal to the GW emission decay time \citep{Hopman+05}
\begin{equation}
    a_c=r\left(\frac{85}{3072} \right)^{1/(3-\alpha)} \left[ \frac{1}{r} \left(\frac{G m_{\rm BH}  t_{\rm relx} }{\sqrt{G M_\bullet}} \right)^{2/3} \right]^{3/(6-2\alpha)} \ ,
\end{equation}
where for consistency we evaluate this critical value at $r_{\rm max}$ (as well as $t_{\rm relx})$, but in the literature this and the rate from Eq.~(\ref{eq:RateRelx}) are evaluated at the sphere of influence. Regardless of the distance we choose (i.e., either $r_{\rm max}$, or the sphere of influence), the  rate depends on the SMBH mass weakly: $ \Gamma_{\rm relx}\approx m_1^{-1/4}$, \citep[e.g.,][]{Hopman+05}. In Figure \ref{fig:EMRIRate}, we show this rate for $r_{\rm max}$ (the sphere of influence), dashed (solid) line.

\section{Discussion}\label{sec:Diss} 
EMRIs are the result of an SMBH that captures a stellar-mass compact object, such as BH. Thus, these are some of the promising GW signals for low-frequency GW detectors such as LISA. Different channels have been suggested to form EMRIs. In particular, two-body relaxation has been proposed as one of the likely physical processes to form EMRIs efficiently. In this process, weak two-body kicks from the population of stars and compact object that surrounds the SMBH can change the BH's orbit over time, {driving} it into the SMBH.  On the other hand, perturbations from SMBH companions, via the EKL mechanism, can excite the SMBH to high eccentricities, thereby forming EMRIs. Here we demonstrated that EMRIs are naturally formed in SMBH binaries with higher efficiency than either of these processes considered alone.

In the presence of an SMBH companion, the EKL mechanism can excite the BHs eccentricity to high values. However, the EKL mechanism's efficiency depends to some extent on the initial conditions \citep[e.g.,][]{Li+14}. Therefore, the small kicks due to two-body relaxation do not need to accumulate to change the angular momentum by order of itself. Instead, they can change the orbital parameters of the stellar mass BH, such as eccentricity, semi-major axis, and argument of periapsis, rendering it in a favorable EKL regime. We show an example of such as system in Figure \ref{fig:TimeEvolution}. Even if the two-body relaxation timescale is orders of magnitude longer than the EKL timescale (see Figure \ref{fig:Cartoon}), the small-kicks are effective as long as they result in a change of angular momentum comparable to that due to EKL. In particular, we suggest that $h/\Delta h|_{\rm relx}$ needs to be within a couple of orders of magnitude (or close to) $h/\Delta h|_{\rm EKL}$. If  $h/\Delta h|_{\rm relx} >>h/\Delta h|_{\rm EKL}$, the angular momentum change $\Delta h$ due to the two-body relaxation can be neglected (see for example Figure \ref{fig:TimeEvolution}). In Figure \ref{fig:Cartoon} we highlight the proposed comparison between the two-body relaxation process and EKL, using $h/\Delta h$ rather than timescales. 

In general, other collective processes may also be considered. For example, resonant relaxations \citep{Rauch+96}, which arise from orbit-averaged mass distribution of the objects around the primary, can be added as well \citep[e.g.,][]{Eilon+09,Kocsis+11,Sridhar+16,Touma+19}. However, scalar and vector resonant relaxation processes modify the angular momentum $\Delta h / h_{\rm RR} \sim t_{\rm res}/P_\bullet$, thus using their timescales to estimate their contribution may not be as misleading as the aforementioned timescale analysis of the two-body relaxation (instead of using $\Delta h / h$). Vector resonant relaxation processes have been added recently to the EKL context and were shown to drive low-inclination configurations to a more EKL favorable regime \citep[e.g.,][]{Hamers+18}. However, the latter study concluded that overall, the combined effect is not very efficient in the context of BH-BH mergers. In contrast, as highlighted here, two-body relaxation results in populating EKL-favorable regimes very efficiently. 

As a proof of concept, we choose a fiducial system composed of an SMBH binary ($m_1=10^7$~M$_\odot$, and $m_1=10^9$~M$_\odot$) on an eccentric orbit $e_{\rm bin}=0.7$, at $1$~pc separation. We begin by considering the effect of the EKL mechanism on stellar black holes around $m_1$ (Figure \ref{fig:MonteCarlo} top panel). Note that all runs include the $1$pN contribution to the inner and outer orbits\footnote{Note that we do not include crossing terms \citep[e.g.,][]{Naoz+12GR,Lim+20}, because their overall effect should be minimal in this configuration. }. Stellar-mass BHs whose pericenter distance passed a critical value are considered as EMRIs.  As noted in previous studies, the efficacy of this mechanism is about $30\%$ \citep[e.g.,][]{NaozSilk14}.  We then systematically add two-body relaxation (middle panel in Figure \ref{fig:MonteCarlo}) and gravitational wave emission (bottom panel). As a result, the efficacy increased to $50-100\%$, meaning nearly all of the stellar mass BHs ended up {descending} into the SMBH, thereby {possibly forming EMRIs,  within a few$\times 10^8$~yr, after a single star formation burst, i.e., not including replenishment. }.   

To highlight the efficiency of this scenario, we extrapolate the EMRI formation rate to different SMBHs. Since EMRIs rate is highly uncertain and is sensitive to the number of BHs as a function of time, we used the $M-\sigma$ relation. Moreover, we rescale our fiducial example by keeping the quadrupole-level of the EKL approximation constant. This means a constant power law and a constant mass ratio and the SMBH binary separation varies accordingly, for example, for $m_1=10^7$~M$_\odot$ ($m_1=10^8$~M$_\odot$), $a_{\rm bin}=1$~pc$ (a_{\rm bin}=2.2$~pc). Furthermore, the number of BHs inside a sphere at which $\epsilon\leq 0.1$ varies accordingly, for $m_1=10^7$~M$_\odot$ ($m_1=10^8$~M$_\odot$), $N_{\rm BH}\approx 331$ ($N_{\rm BH}\sim 1979$). We depict the rescaling in Figure \ref{fig:Acc}.

Even for this simple scaling, it is clear that having the entire population of BHs, (or even just $50\%$) becoming EMRIs has large implications on the EMRI rate. We compare the predicted EMRI rate from this scenario to the prediction from two-body relaxation only in Figure \ref{fig:EMRIRate}. As depicted in this Figure, the EMRI rate in SMBH binaries is orders of magnitude larger than in isolated SMBHs. Additionally, the dependency on the SMBH mass is different, offering a potential way to disentangle between the different scenarios. Furthermore, because SMBH binaries are expected to be ubiquitous in the Universe, our results suggest that the EMRI rate may be much higher than nominal estimations. In particular, post star burst galaxies may be interesting candidates for enhanced EMRIs formation as they possibly host a SMBH binary.  Moreover, this result suggests that the observed EMRI rate may be used to constrain the prevalence of SMBH binaries in the Universe.  


\acknowledgments
 We  thank the referee for useful comments. SN acknowledges the partial support from NASA ATP  80NSSC20K0505  and thanks Howard and Astrid Preston for their generous support. SR thanks the Nina Byers Fellowship, the Charles E Young Fellowship, and the Michael A. Jura Memorial Graduate Award for support, as well as partial support  from NASA ATP  80NSSC20K0505. EM acknowledges the support of  s Howard and Astrid Preston, the   Mani  L.  Bhaumik  Institute  for  Theoretical  Physics, and  as well as partial support  from NASA ATP  80NSSC20K0505. DM acknowledges the partial support from NSF graduate fellowship, the Eugene Cota-Robles Fellowship, and the NASA ATP  80NSSC20K0505. B.M. is grateful for the AAUW American Fellowship, and the UCSC Presidents Dissertation Fellowship. E.R.-R. and B.M. are grateful for support from the Packard Foundation, Heising-Simons Foundation, NSF (AST-1615881, AST-1911206 and AST-1852393), Swift (80NSSC21K1409, 80NSSC19K1391) and Chandra (GO9-20122X).

\appendix
\section{A: The post kick orbital parameters} \label{sec:app}
Consider a BH orbiting SMBH. In the plane of the ellipse we can define the separation vector as ${\bf r}_\bullet = r_\bullet (\cos f_\bullet,\sin f_\bullet ,0)$, where $f_\bullet$ is the true anomaly and  
\begin{equation}
    r_\bullet=\frac{a_\bullet(1-e_\bullet^2)}{1+e_\bullet \cos f_\bullet} \ .
\end{equation}
The associated velocity vector at the plane of the ellipse is: ${\bf v}_\bullet = h / a_\bullet(-\sin f_\bullet, e_\bullet+ \cos f_\bullet,0)/( 1-e_\bullet^2)$. These vectors are projected onto the invariable plane, where in the case of test-particle EKL is simply the plane of the outer orbit \citep[e.g.,][]{LN}. 
Thus, we rotate the the separation and velocity vectors at each time-step given their argument of perihapsis, $\omega$, longitude of ascending nodes, $\Omega$ and inclination $i$. For example, and similarly for the velocity vector, we have:
\begin{equation}
{\bf r}_{\bullet,\textnormal{inv}}=R_z(\Omega)R_x(i)R_z(\omega){\bf r}_{\bullet,\textnormal{ell}} \ ,
\end{equation}
where the subscript ``inv'' and ``ell'' refer to the invariable and
ellipse coordinate systems, respectively.  Given a rotation angle $\theta$, the rotation matrices $R_z$ and $R_x$  are
\begin{equation}
R_z(\theta) = \left(
\begin{array}{ccc}
\cos\theta & -\sin\theta & 0 \\
\sin\theta & \cos\theta & 0 \\
0 & 0 & 1
\end{array}\right)
\end{equation}
and
\begin{equation}
R_x(\theta) = \left(
\begin{array}{ccc}
1 & 0 & 0 \\
0 &\cos\theta & -\sin\theta \\
0 &\sin\theta & \cos\theta
\end{array}\right) \ .
\end{equation}

A two-body encounter can change its velocity by:
\begin{equation}
    \Delta v = v_\bullet\sqrt{ \frac{P_\bullet}{t_{relx}}} \ . 
\end{equation}
We model this change as a random walk, applying a single isotropic, {\it instantaneous}, kick to the BH velocity once per $P_\bullet$. Each directional component of this 3D kick is drawn from a Gaussian distribution with a zero average and a standard deviation of $\Delta v_j/\sqrt{3}$, where $j$ is $1,2,3$ for the three components of the velocity vector. The  instantaneous assumption means that ${r}_\bullet$ is kept constant during the kick \citep[see][]{Kalogera00}. 

Thus, post-kick, the new velocity vector (in the invariable plane) is: 
${\bf v}_{\bullet,p} = {\bf \Delta v} + {\bf v}_{\bullet}$, where we dropped the subscript ``inv'' to avoid clutter and the subscript ``p'' means post-kick. The angular momentum post-kick is ${\bf h}_p = {\bf r}_\bullet \times {\bf v}_{\bullet,p}$. Thus, it is straightforward to find the orbital parameters. Specifically, the semi-major axis of the BH post-kick is:
\begin{equation}
    a_{\bullet,p} = \left( \frac{2}{r_\bullet} - \frac{v_{\bullet,p}^2}{G m_1} \right)^{-1} \ ,
\end{equation}
the post-kick eccentricity is:
\begin{equation}
    e_{\bullet,p} = \sqrt{1-\frac{h_p^2}{G m_1 a_{\bullet,p}} }\ .
\end{equation}
Because the $z$ axis is defined by the outer orbit, the new inclination is $\cos i_p = h_{p,z}/h_p$, where $ h_{p,z}$ is the $z$ component of the post-kick angular momentum. 

 The post-kick longitude of acsending nodes is:
\begin{equation}
    \Omega_p = \arctan_2 \left(\pm\frac{h_{p,1}}{h_p\sin i_p}, \mp  \frac{h_{p,2}}{h_p\sin i_p}\right) \ .
\end{equation}
The post-kick true anomaly is: 
\begin{equation}
    f_p = \arctan_2 \left(\frac{a_{\bullet,p} (1-e_{\bullet,p}^2)}{h_p e_{\bullet,p}}\dot{R}, \frac{1}{e_{\bullet,p}}\bigg[ \frac{a_{\bullet,p}(1-e_{\bullet,p}^2)}{r_\bullet} -1 \bigg]\right) \ ,
\end{equation}
where  $\dot{R} = \pm \sqrt{v^2_{\bullet,p}- h_p^2/r_\bullet^2}$, where the sign is defined by the sign of ${\bf r}_\bullet \cdot {\bf v}_{\bullet,p}$ \citep[e.g.][]{MD00}. The post-kick argument of pericenter is then:
\begin{eqnarray}
    \omega_p &=&  \arctan_2 \left(\frac{r_{\bullet,3}}{r_\bullet\sin i_p},\bigg[ \frac{r_{\bullet,1}}{r_\bullet} +  \frac{ r_{\bullet,3}\sin \omega_p \cos i_p }{r_\bullet\sin i_p} \bigg] \sec \Omega_p \right) - f_p
\end{eqnarray}

\section{B: Plunging orbits and an Example of Signal to Noise in the LISA band}\label{sec:SNR}

{We first differentiate between plunging orbits and EMRIs, where the former is described as a burst associated with their peri-center passage. Our adopted stopping condition of  $R_{\rm sch}=8 Gm_1/c^2$ means that beyond this threshold the BH trajectory will be modified by Kerr geometry and special relativity \citep[e.g.,][]{Schnittman15,Schnittman+18,Yunes+08,Berry+13}. The specific trajectories are beyond the scope of this study. Nonetheless, in the presence of GW emission, we can roughly estimate the fraction of systems that are more likely to appear as EMRIs rather than GW bursts. For that, we first confirmed that all of the systems in the EKL (+GR) + two-body relaxation indeed reach the Schwarzschild radius by integrating all the systems below the solid line in the bottom panel of Figure \ref{fig:MonteCarlo}. 

Second, examining the integration prior to the threshold we found that $\sim 40\%$ of the system reach a configuration for which $P_\bullet\leq 10$~yr, and $a_\bullet(1-e_\bullet)<1$~au. This specific configuration is chosen such that the characteristic strain will appear in the LISA band, resulting in mHz signals (see below). Assuming LISA lifetime to be about $10$~yr. We emphasize that the $40\%$ estimation is rather conservative because, as mentioned, even the plunged BHs trajectories may spend a long time zooming in the SMBH's ergosphere, where GW emission may alter their separation can result in an EMRI-like signal. Note that even when the BH period is smaller than $10$~years (roughly equivalent to S0-2's orbital period), two-body relaxation may still result in small kicks, about $0.0003$ of the BH velocity, according to Eq.~(\ref{eq:Deltav}).  Thus, overall the orbit will not substantially change over the BH period\footnote{Note that we are not taking into account star-BH collisions and tidal interactions that may result in electromagnetic signatures or larger BHs \citep[e.g.,][]{Metzger+21,Rose+22,Kremer+22}.} .        

}

\begin{figure}
  \begin{center} 
    \includegraphics[width=0.6\linewidth]{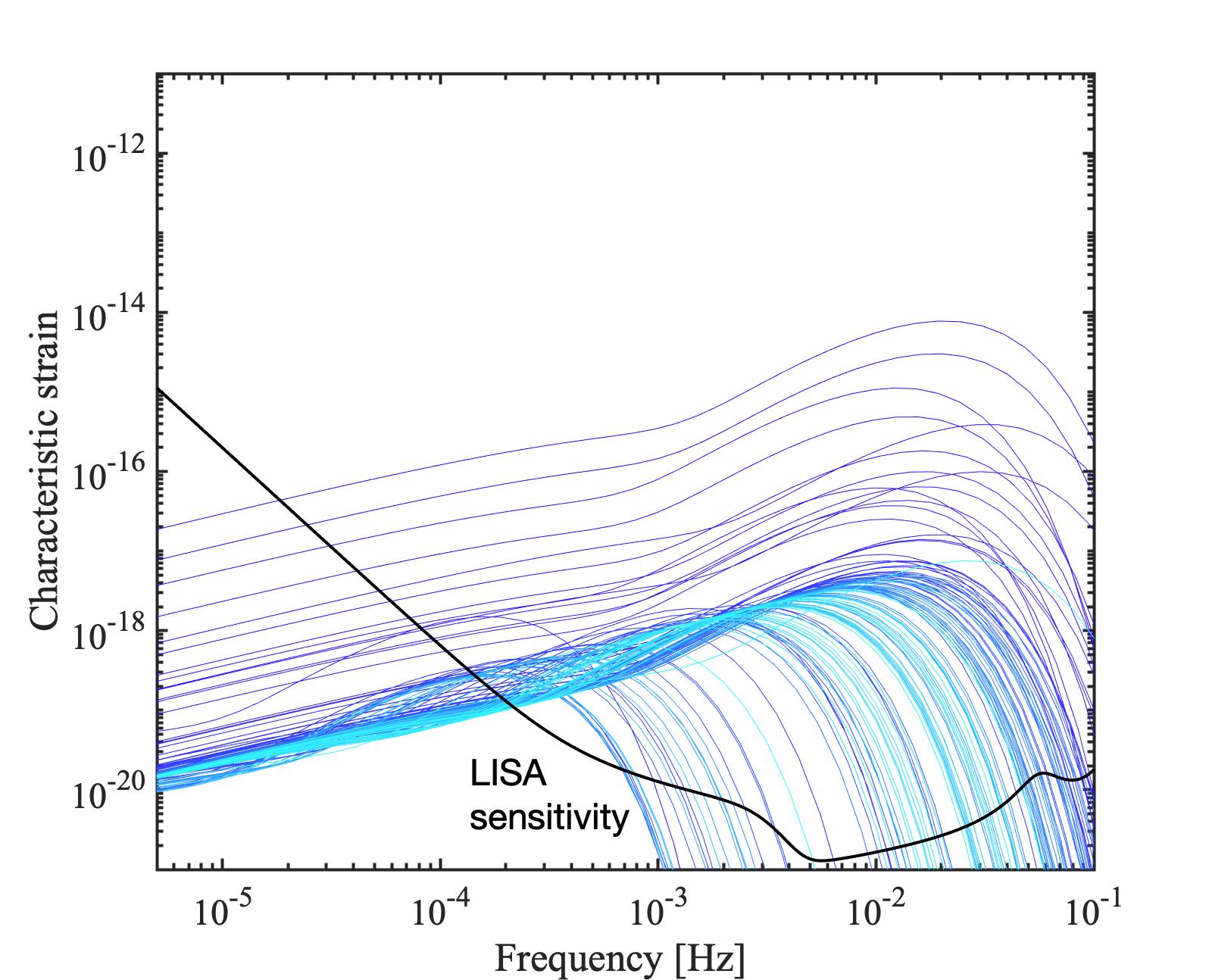}
  \end{center} 
  \caption{  \upshape {\bf An example of the characteristic strain for $150$ (chosen randomly) of the runs that reached $P_\bullet\leq 10$~yr, and $a_\bullet(1-e_\bullet)<1$~au. } We consider the case which includes EKL (+GR) + two-body relaxation + GW. 
  } \label{fig:LISA} 
\end{figure}

To estimate the signal to noise we follow \citet{Robson+18,Robson+19}. The strain and thus the SNR depend on the the orbital period, the eccentricity, and the luminosity distance. As a proof of concept we depict in Figure \ref{fig:LISA} the characteristic strain for all of the runs that crossed $R_{\rm sch}$ in our nominal system (i.e., all the point below the line in Figure \ref{fig:MonteCarlo}). For this example, we adopt a luminosity distance of $0.7$~Mpc, and LISA observation time of $10$~years.  We find that $52\%$ of the systems have a SNR $>5$. Out of these systems $3\%$ have GW dissipation timescale which is shorter than $10$~years, which implies that a more careful analysis of the  characteristic strain should be conducted for them  \citep[e.g.,][]{Barack+04}. Eccentricity oscillations  due to the EKL signature on the  characteristic strain \citep[e.g.,][]{Hoang+19,Deme+20} are unlikely to be detected in this configuration.  

\vspace{0.2cm}
\bibliographystyle{hapj}
\bibliography{Binary}

\begin{thebibliography}{109}
\expandafter\ifx\csname natexlab\endcsname\relax\def\natexlab#1{#1}\fi

\bibitem[{{Aharon} \& {Perets}(2016)}]{Aharon+16}
{Aharon}, D., \& {Perets}, H.~B. 2016, \apjl, 830, L1, 1609.01715

\bibitem[{{Alexander} \& {Hopman}(2009)}]{Alexander+09}
{Alexander}, T., \& {Hopman}, C. 2009, \apj, 697, 1861, 0808.3150

\bibitem[{{Amaro-Seoane}(2018)}]{Amaro18}
{Amaro-Seoane}, P. 2018, Living Reviews in Relativity, 21, 4, 1205.5240

\bibitem[{{Amaro-Seoane} {et~al.}(2017){Amaro-Seoane}, {Audley}, {Babak},
  {Baker}, {Barausse}, {Bender}, {Berti}, {Binetruy}, {Born}, {Bortoluzzi},
  {Camp}, {Caprini}, {Cardoso}, {Colpi}, {Conklin}, {Cornish}, {Cutler},
  {Danzmann}, {Dolesi}, {Ferraioli}, {Ferroni}, {Fitzsimons}, {Gair}, {Gesa
  Bote}, {Giardini}, {Gibert}, {Grimani}, {Halloin}, {Heinzel}, {Hertog},
  {Hewitson}, {Holley-Bockelmann}, {Hollington}, {Hueller}, {Inchauspe},
  {Jetzer}, {Karnesis}, {Killow}, {Klein}, {Klipstein}, {Korsakova}, {Larson},
  {Livas}, {Lloro}, {Man}, {Mance}, {Martino}, {Mateos}, {McKenzie},
  {McWilliams}, {Miller}, {Mueller}, {Nardini}, {Nelemans}, {Nofrarias},
  {Petiteau}, {Pivato}, {Plagnol}, {Porter}, {Reiche}, {Robertson},
  {Robertson}, {Rossi}, {Russano}, {Schutz}, {Sesana}, {Shoemaker}, {Slutsky},
  {Sopuerta}, {Sumner}, {Tamanini}, {Thorpe}, {Troebs}, {Vallisneri},
  {Vecchio}, {Vetrugno}, {Vitale}, {Volonteri}, {Wanner}, {Ward}, {Wass},
  {Weber}, {Ziemer}, \& {Zweifel}}]{Amaro-Seoane+17}
{Amaro-Seoane}, P. {et~al.} 2017, arXiv e-prints, arXiv:1702.00786, 1702.00786

\bibitem[{{Amaro-Seoane} \& {Preto}(2011)}]{Amaro+11}
{Amaro-Seoane}, P., \& {Preto}, M. 2011, Classical and Quantum Gravity, 28,
  094017, 1010.5781

\bibitem[{{Antognini}(2015)}]{Antognini15}
{Antognini}, J.~M.~O. 2015, \mnras, 452, 3610, 1504.05957

\bibitem[{{Babak} {et~al.}(2017){Babak}, {Gair}, {Sesana}, {Barausse},
  {Sopuerta}, {Berry}, {Berti}, {Amaro-Seoane}, {Petiteau}, \&
  {Klein}}]{Babak+17}
{Babak}, S. {et~al.} 2017, \prd, 95, 103012, 1703.09722

\bibitem[{{Bahcall} \& {Wolf}(1976)}]{Bahcall+76}
{Bahcall}, J.~N., \& {Wolf}, R.~A. 1976, \apj, 209, 214

\bibitem[{{Baker} {et~al.}(2019){Baker}, {Bellovary}, {Bender}, {Berti},
  {Caldwell}, {Camp}, {Conklin}, {Cornish}, {Cutler}, {DeRosa}, {Eracleous},
  {Ferrara}, {Francis}, {Hewitson}, {Holley-Bockelmann}, {Hornschemeier},
  {Hogan}, {Kamai}, {Kelly}, {Shapiro Key}, {Larson}, {Livas},
  {Manthripragada}, {McKenzie}, {McWilliams}, {Mueller}, {Natarajan}, {Numata},
  {Rioux}, {Sankar}, {Schnittman}, {Shoemaker}, {Shoemaker}, {Slutsky},
  {Spero}, {Stebbins}, {Thorpe}, {Vallisneri}, {Ware}, {Wass}, {Yu}, \&
  {Ziemer}}]{Baker+19}
{Baker}, J. {et~al.} 2019, arXiv e-prints, arXiv:1907.06482, 1907.06482

\bibitem[{{Bar-Or} \& {Alexander}(2016)}]{Bar-Or+16}
{Bar-Or}, B., \& {Alexander}, T. 2016, \apj, 820, 129, 1508.01390

\bibitem[{{Barack} \& {Cutler}(2004)}]{Barack+04}
{Barack}, L., \& {Cutler}, C. 2004, \prd, 69, 082005, gr-qc/0310125

\bibitem[{{Batcheldor} {et~al.}(2010){Batcheldor}, {Robinson}, {Axon},
  {Perlman}, \& {Merritt}}]{Batcheldor+10}
{Batcheldor}, D., {Robinson}, A., {Axon}, D.~J., {Perlman}, E.~S., \&
  {Merritt}, D. 2010, \apjl, 717, L6, 1005.2173

\bibitem[{{Berry} \& {Gair}(2013)}]{Berry+13}
{Berry}, C.~P.~L., \& {Gair}, J.~R. 2013, \mnras, 433, 3572, 1306.0774

\bibitem[{{Bhaskar} {et~al.}(2021){Bhaskar}, {Li}, {Hadden}, {Payne}, \&
  {Holman}}]{Bhaskar+21}
{Bhaskar}, H., {Li}, G., {Hadden}, S., {Payne}, M.~J., \& {Holman}, M.~J. 2021,
  \aj, 161, 48, 2008.04335

\bibitem[{{Bianchi} {et~al.}(2008){Bianchi}, {Chiaberge}, {Piconcelli},
  {Guainazzi}, \& {Matt}}]{Bianchi+08}
{Bianchi}, S., {Chiaberge}, M., {Piconcelli}, E., {Guainazzi}, M., \& {Matt},
  G. 2008, MNRAS, 386, 105, 0802.0825

\bibitem[{{Binney} \& {Tremaine}(2008)}]{Binney+TremaineBook}
{Binney}, J., \& {Tremaine}, S. 2008, {Galactic Dynamics: Second Edition}

\bibitem[{{Blaes} {et~al.}(2002){Blaes}, {Lee}, \& {Socrates}}]{Bla+02}
{Blaes}, O., {Lee}, M.~H., \& {Socrates}, A. 2002, \apj, 578, 775,
  arXiv:astro-ph/0203370

\bibitem[{{Bode} \& {Wegg}(2014)}]{Bode+14}
{Bode}, J.~N., \& {Wegg}, C. 2014, \mnras, 438, 573

\bibitem[{{Bogdanovi{\'c}} {et~al.}(2009){Bogdanovi{\'c}}, {Eracleous}, \&
  {Sigurdsson}}]{Bogdanovic+09}
{Bogdanovi{\'c}}, T., {Eracleous}, M., \& {Sigurdsson}, S. 2009, \apj, 697,
  288, 0809.3262

\bibitem[{{Boroson} \& {Lauer}(2009)}]{Boroson+09}
{Boroson}, T.~A., \& {Lauer}, T.~R. 2009, \nat, 458, 53, 0901.3779

\bibitem[{{Bradnick} {et~al.}(2017){Bradnick}, {Mandel}, \&
  {Levin}}]{Bradnick+17}
{Bradnick}, B., {Mandel}, I., \& {Levin}, Y. 2017, \mnras, 469, 2042,
  1703.05796

\bibitem[{{Callegari} {et~al.}(2009){Callegari}, {Mayer}, {Kazantzidis},
  {Colpi}, {Governato}, {Quinn}, \& {Wadsley}}]{Callegari+09}
{Callegari}, S., {Mayer}, L., {Kazantzidis}, S., {Colpi}, M., {Governato}, F.,
  {Quinn}, T., \& {Wadsley}, J. 2009, ApJ-Lett, 696, L89, 0811.0615

\bibitem[{{Chen} \& {Han}(2018)}]{chen+18}
{Chen}, X., \& {Han}, W.-B. 2018, Communications Physics, 1, 53, 1801.05780

\bibitem[{{Chen} \& {Liu}(2013)}]{Chen+13}
{Chen}, X., \& {Liu}, F.~K. 2013, \apj, 762, 95, 1211.4609

\bibitem[{{Chen} {et~al.}(2008){Chen}, {Liu}, \& {Magorrian}}]{Chen+08}
{Chen}, X., {Liu}, F.~K., \& {Magorrian}, J. 2008, \apj, 676, 54, 0712.0246

\bibitem[{{Chen} {et~al.}(2009){Chen}, {Madau}, {Sesana}, \& {Liu}}]{Chen+09}
{Chen}, X., {Madau}, P., {Sesana}, A., \& {Liu}, F.~K. 2009, \apjl, 697, L149,
  0904.4481

\bibitem[{{Chen} {et~al.}(2011){Chen}, {Sesana}, {Madau}, \& {Liu}}]{Chen+11}
{Chen}, X., {Sesana}, A., {Madau}, P., \& {Liu}, F.~K. 2011, \apj, 729, 13,
  1012.4466

\bibitem[{{Comerford} {et~al.}(2009){Comerford}, {Griffith}, {Gerke}, {Cooper},
  {Newman}, {Davis}, \& {Stern}}]{Comerford+09bin}
{Comerford}, J.~M., {Griffith}, R.~L., {Gerke}, B.~F., {Cooper}, M.~C.,
  {Newman}, J.~A., {Davis}, M., \& {Stern}, D. 2009, ApJ-Lett, 702, L82,
  0906.3517

\bibitem[{{Comerford} {et~al.}(2018){Comerford}, {Nevin}, {Stemo},
  {M{\"u}ller-S{\'a}nchez}, {Barrows}, {Cooper}, \& {Newman}}]{Comerford+18}
{Comerford}, J.~M., {Nevin}, R., {Stemo}, A., {M{\"u}ller-S{\'a}nchez}, F.,
  {Barrows}, R.~S., {Cooper}, M.~C., \& {Newman}, J.~A. 2018, \apj, 867, 66,
  1810.11543

\bibitem[{{Deane} {et~al.}(2014){Deane}, {Paragi}, {Jarvis}, {Coriat},
  {Bernardi}, {Fender}, {Frey}, {Heywood}, {Kl{\"o}ckner}, {Grainge}, \&
  {Rumsey}}]{Deane+14}
{Deane}, R.~P. {et~al.} 2014, \nat, 511, 57, 1406.6365

\bibitem[{{Deme} {et~al.}(2020){Deme}, {Hoang}, {Naoz}, \& {Kocsis}}]{Deme+20}
{Deme}, B., {Hoang}, B.-M., {Naoz}, S., \& {Kocsis}, B. 2020, \apj, 901, 125,
  2005.03677

\bibitem[{{Di Matteo} {et~al.}(2005){Di Matteo}, {Springel}, \&
  {Hernquist}}]{DiMatteo+05}
{Di Matteo}, T., {Springel}, V., \& {Hernquist}, L. 2005, \nat, 433, 604,
  astro-ph/0502199

\bibitem[{{Dotti} {et~al.}(2009){Dotti}, {Montuori}, {Decarli}, {Volonteri},
  {Colpi}, \& {Haardt}}]{Dotti+09}
{Dotti}, M., {Montuori}, C., {Decarli}, R., {Volonteri}, M., {Colpi}, M., \&
  {Haardt}, F. 2009, MNRAS, 398, L73, 0809.3446

\bibitem[{{Eilon} {et~al.}(2009){Eilon}, {Kupi}, \& {Alexander}}]{Eilon+09}
{Eilon}, E., {Kupi}, G., \& {Alexander}, T. 2009, \apj, 698, 641, 0807.1430

\bibitem[{{Ford} {et~al.}(2000){Ford}, {Joshi}, {Rasio}, \&
  {Zbarsky}}]{Ford00Pls}
{Ford}, E.~B., {Joshi}, K.~J., {Rasio}, F.~A., \& {Zbarsky}, B. 2000, \apj,
  528, 336, arXiv:astro-ph/9905347

\bibitem[{{Fragione} {et~al.}(2020){Fragione}, {Loeb}, {Kremer}, \&
  {Rasio}}]{Fragione+20}
{Fragione}, G., {Loeb}, A., {Kremer}, K., \& {Rasio}, F.~A. 2020, \apj, 897,
  46, 2002.02975

\bibitem[{{Freitag}(2001)}]{Freitag01}
{Freitag}, M. 2001, Classical and Quantum Gravity, 18, 4033, astro-ph/0107193

\bibitem[{{Generozov} \& {Madigan}(2020)}]{Generozov+20}
{Generozov}, A., \& {Madigan}, A.-M. 2020, \apj, 896, 137, 2002.10547

\bibitem[{{GRAVITY Collaboration} {et~al.}(2020){GRAVITY Collaboration},
  {Abuter}, {Amorim}, {Baub{\"o}ck}, {Berger}, {Bonnet}, {Brand ner},
  {Cardoso}, {Cl{\'e}net}, {de Zeeuw}, {Dexter}, {Eckart}, {Eisenhauer},
  {F{\"o}rster Schreiber}, {Garcia}, {Gao}, {Gendron}, {Genzel}, {Gillessen},
  {Habibi}, {Haubois}, {Henning}, {Hippler}, {Horrobin}, {Jim{\'e}nez-Rosales},
  {Jochum}, {Jocou}, {Kaufer}, {Kervella}, {Lacour}, {Lapeyr{\`e}re}, {Le
  Bouquin}, {L{\'e}na}, {Nowak}, {Ott}, {Paumard}, {Perraut}, {Perrin},
  {Pfuhl}, {Rodr{\'\i}guez-Coira}, {Shangguan}, {Scheithauer}, {Stadler},
  {Straub}, {Straubmeier}, {Sturm}, {Tacconi}, {Vincent}, {von Fellenberg},
  {Waisberg}, {Widmann}, {Wieprecht}, {Wiezorrek}, {Woillez}, {Yazici}, \&
  {Zins}}]{Gravity+20}
{GRAVITY Collaboration} {et~al.} 2020, \aap, 636, L5, 2004.07187

\bibitem[{{Green} {et~al.}(2010){Green}, {Myers}, {Barkhouse}, {Mulchaey},
  {Bennert}, {Cox}, \& {Aldcroft}}]{Green+10}
{Green}, P.~J., {Myers}, A.~D., {Barkhouse}, W.~A., {Mulchaey}, J.~S.,
  {Bennert}, V.~N., {Cox}, T.~J., \& {Aldcroft}, T.~L. 2010, \apj, 710, 1578,
  1001.1738

\bibitem[{{Gualandris} \& {Merritt}(2009)}]{Gualandris+09}
{Gualandris}, A., \& {Merritt}, D. 2009, \apj, 705, 361, 0905.4514

\bibitem[{{G{\"u}rkan} \& {Rasio}(2005)}]{Grkan+20}
{G{\"u}rkan}, M.~A., \& {Rasio}, F.~A. 2005, \apj, 628, 236, astro-ph/0412452

\bibitem[{{Hamers} {et~al.}(2018){Hamers}, {Bar-Or}, {Petrovich}, \&
  {Antonini}}]{Hamers+18}
{Hamers}, A.~S., {Bar-Or}, B., {Petrovich}, C., \& {Antonini}, F. 2018, \apj,
  865, 2, 1805.10313

\bibitem[{{Hansen} \& {Milosavljevi{\'c}}(2003)}]{Hansen+03}
{Hansen}, B.~M.~S., \& {Milosavljevi{\'c}}, M. 2003, ApJ-Lett, 593, L77,
  arXiv:astro-ph/0306074

\bibitem[{{Hansen} \& {Naoz}(2020)}]{Hansen+20}
{Hansen}, B. M.~S., \& {Naoz}, S. 2020, \mnras, 499, 1682, 2011.07103

\bibitem[{{Haster} {et~al.}(2016){Haster}, {Antonini}, {Kalogera}, \&
  {Mandel}}]{Haster+16}
{Haster}, C.-J., {Antonini}, F., {Kalogera}, V., \& {Mandel}, I. 2016, \apj,
  832, 192, 1606.07097

\bibitem[{{Hoang} {et~al.}(2019){Hoang}, {Naoz}, {Kocsis}, {Farr}, \&
  {McIver}}]{Hoang+19}
{Hoang}, B.-M., {Naoz}, S., {Kocsis}, B., {Farr}, W.~M., \& {McIver}, J. 2019,
  \apj, 875, L31, 1903.00134

\bibitem[{{Hopkins} {et~al.}(2006){Hopkins}, {Hernquist}, {Cox}, {Di Matteo},
  {Robertson}, \& {Springel}}]{Hopkins+06}
{Hopkins}, P.~F., {Hernquist}, L., {Cox}, T.~J., {Di Matteo}, T., {Robertson},
  B., \& {Springel}, V. 2006, ApJS, 163, 1, astro-ph/0506398

\bibitem[{{Hopman}(2009)}]{Hopman09}
{Hopman}, C. 2009, Classical and Quantum Gravity, 26, 094028, 0901.1667

\bibitem[{{Hopman} \& {Alexander}(2005)}]{Hopman+05}
{Hopman}, C., \& {Alexander}, T. 2005, \apj, 629, 362, astro-ph/0503672

\bibitem[{{Hopman} \& {Alexander}(2006)}]{Hopman+06seg}
------. 2006, \apjl, 645, L133, astro-ph/0603324

\bibitem[{{Iwasa} \& {Seto}(2016)}]{Iwasa+16}
{Iwasa}, M., \& {Seto}, N. 2016, \prd, 93, 124024, 1508.05762

\bibitem[{{Kalogera}(2000)}]{Kalogera00}
{Kalogera}, V. 2000, \apj, 541, 319, astro-ph/9911417

\bibitem[{{Kocsis} \& {Tremaine}(2011)}]{Kocsis+11}
{Kocsis}, B., \& {Tremaine}, S. 2011, \mnras, 412, 187, 1006.0001

\bibitem[{{Komossa} {et~al.}(2003){Komossa}, {Burwitz}, {Hasinger}, {Predehl},
  {Kaastra}, \& {Ikebe}}]{Komossa+03}
{Komossa}, S., {Burwitz}, V., {Hasinger}, G., {Predehl}, P., {Kaastra}, J.~S.,
  \& {Ikebe}, Y. 2003, ApJ-Lett, 582, L15, astro-ph/0212099

\bibitem[{{Komossa} {et~al.}(2008){Komossa}, {Zhou}, \& {Lu}}]{Komossa+08}
{Komossa}, S., {Zhou}, H., \& {Lu}, H. 2008, ApJ-Lett, 678, L81, 0804.4585

\bibitem[{{Kozai}(1962)}]{Kozai}
{Kozai}, Y. 1962, \aj, 67, 591

\bibitem[{{Kremer} {et~al.}(2022){Kremer}, {Lombardi}, {Lu}, {Piro}, \&
  {Rasio}}]{Kremer+22}
{Kremer}, K., {Lombardi}, James~C., J., {Lu}, W., {Piro}, A.~L., \& {Rasio},
  F.~A. 2022, arXiv e-prints, arXiv:2201.12368, 2201.12368

\bibitem[{{Li} {et~al.}(2014{\natexlab{a}}){Li}, {Naoz}, {Holman}, \&
  {Loeb}}]{Li+14}
{Li}, G., {Naoz}, S., {Holman}, M., \& {Loeb}, A. 2014{\natexlab{a}}, \apj,
  791, 86, 1405.0494

\bibitem[{{Li} {et~al.}(2014{\natexlab{b}}){Li}, {Naoz}, {Kocsis}, \&
  {Loeb}}]{Li+13}
{Li}, G., {Naoz}, S., {Kocsis}, B., \& {Loeb}, A. 2014{\natexlab{b}}, \apj,
  785, 116, 1310.6044

\bibitem[{{Li} {et~al.}(2015){Li}, {Naoz}, {Kocsis}, \& {Loeb}}]{Li+15}
------. 2015, \mnras, 451, 1341, 1502.03825

\bibitem[{{Li} {et~al.}(2020){Li}, {Bogdanovi{\'c}}, \&
  {Ballantyne}}]{Li+20Pair}
{Li}, K., {Bogdanovi{\'c}}, T., \& {Ballantyne}, D.~R. 2020, \apj, 896, 113,
  2006.08520

\bibitem[{{Lidov}(1962)}]{Lidov}
{Lidov}, M.~L. 1962, planss, 9, 719

\bibitem[{{Lim} \& {Rodriguez}(2020)}]{Lim+20}
{Lim}, H., \& {Rodriguez}, C.~L. 2020, \prd, 102, 064033, 2001.03654

\bibitem[{{Lithwick} \& {Naoz}(2011{\natexlab{a}})}]{LN11}
{Lithwick}, Y., \& {Naoz}, S. 2011{\natexlab{a}}, \apj, 742, 94, 1106.3329

\bibitem[{{Lithwick} \& {Naoz}(2011{\natexlab{b}})}]{LN}
------. 2011{\natexlab{b}}, \apj, 742, 94, 1106.3329

\bibitem[{{Liu} {et~al.}(2010){Liu}, {Greene}, {Shen}, \&
  {Strauss}}]{Liu+10kpc}
{Liu}, X., {Greene}, J.~E., {Shen}, Y., \& {Strauss}, M.~A. 2010, ApJ-Lett,
  715, L30, 1003.3467

\bibitem[{{Lu} \& {Naoz}(2019)}]{Lu+19}
{Lu}, C.~X., \& {Naoz}, S. 2019, \mnras, 484, 1506, 1805.06897

\bibitem[{{Lu} {et~al.}(2013){Lu}, {Do}, {Ghez}, {Morris}, {Yelda}, \&
  {Matthews}}]{Lu+13}
{Lu}, J.~R., {Do}, T., {Ghez}, A.~M., {Morris}, M.~R., {Yelda}, S., \&
  {Matthews}, K. 2013, \apj, 764, 155, 1301.0540

\bibitem[{{Maillard} {et~al.}(2004){Maillard}, {Paumard}, {Stolovy}, \&
  {Rigaut}}]{Maillard+04}
{Maillard}, J.~P., {Paumard}, T., {Stolovy}, S.~R., \& {Rigaut}, F. 2004, \aap,
  423, 155, arXiv:astro-ph/0404450

\bibitem[{{McConnell} \& {Ma}(2013)}]{McConnell+13}
{McConnell}, N.~J., \& {Ma}, C.-P. 2013, \apj, 764, 184, 1211.2816

\bibitem[{Mei {et~al.}(2020)Mei, Bai, Bao, Barausse, Cai, Canuto, Cao, Chen,
  Chen, Ding, Duan, Fan, Feng, Fu, Gao, Gao, Gong, Gou, Gu, Gu, He, Hendry,
  Hong, Hu, Hu, Hu, Huang, Huang, Jiang, Jiang, Jiang, Jiang, Jin, Korol, Li,
  Li, Li, Li, Li, Li, Li, Li, Li, Liang, Liang, Liao, Liu, Liu, Liu, Liu, Liu,
  Liu, Liu, Lu, Lu, Lu, Luo, Luo, Milyukov, Ming, Pi, Qin, Qu, Sesana, Shao,
  Shi, Su, Tan, Tan, Tan, Tu, Wang, Wang, Wang, Wang, Wang, Wang, Wang, Wang,
  Wang, Wang, Wang, Wei, Wu, Xiao, Xu, Xue, Yang, Yang, Yang, Yang, Ye, Yeh,
  Yu, Zhai, Zhang, Zhang, Zhang, Zhang, Zhang, Zhang, Zhang, Zhou, Zhou, Zhou,
  Zhu, Zi, \& Luo}]{Mei+20}
Mei, J. {et~al.} 2020, Progress of Theoretical and Experimental Physics, 2021,
  https://academic.oup.com/ptep/article-pdf/2021/5/05A107/37953035/ptaa114.pdf,
  05A107

\bibitem[{{Metzger} {et~al.}(2021){Metzger}, {Stone}, \&
  {Gilbaum}}]{Metzger+21}
{Metzger}, B.~D., {Stone}, N.~C., \& {Gilbaum}, S. 2021, arXiv e-prints,
  arXiv:2107.13015, 2107.13015

\bibitem[{{Miller} {et~al.}(2005){Miller}, {Freitag}, {Hamilton}, \&
  {Lauburg}}]{Miller+05}
{Miller}, M.~C., {Freitag}, M., {Hamilton}, D.~P., \& {Lauburg}, V.~M. 2005,
  \apjl, 631, L117, astro-ph/0507133

\bibitem[{{Murray} \& {Dermott}(2000)}]{MD00}
{Murray}, C.~D., \& {Dermott}, S.~F. 2000, {Solar System Dynamics}, ed.
  {Murray, C.~D.~\& Dermott, S.~F.}

\bibitem[{{Naoz}(2016)}]{Naoz16}
{Naoz}, S. 2016, \araa, 54, 441, 1601.07175

\bibitem[{{Naoz} {et~al.}(2013{\natexlab{a}}){Naoz}, {Farr}, {Lithwick},
  {Rasio}, \& {Teyssandier}}]{Naoz+11sec}
{Naoz}, S., {Farr}, W.~M., {Lithwick}, Y., {Rasio}, F.~A., \& {Teyssandier}, J.
  2013{\natexlab{a}}, MNRAS, 431, 2155, 1107.2414

\bibitem[{{Naoz} {et~al.}(2013{\natexlab{b}}){Naoz}, {Kocsis}, {Loeb}, \&
  {Yunes}}]{Naoz+12GR}
{Naoz}, S., {Kocsis}, B., {Loeb}, A., \& {Yunes}, N. 2013{\natexlab{b}}, \apj,
  773, 187, 1206.4316

\bibitem[{{Naoz} {et~al.}(2017){Naoz}, {Li}, {Zanardi}, {de El{\'\i}a}, \& {Di
  Sisto}}]{Naoz+17}
{Naoz}, S., {Li}, G., {Zanardi}, M., {de El{\'\i}a}, G.~C., \& {Di Sisto},
  R.~P. 2017, \aj, 154, 18, 1701.03795

\bibitem[{{Naoz} \& {Silk}(2014)}]{NaozSilk14}
{Naoz}, S., \& {Silk}, J. 2014, \apj, 795, 102, 1409.5432

\bibitem[{{Naoz} {et~al.}(2019){Naoz}, {Silk}, \& {Schnittman}}]{Naoz+19}
{Naoz}, S., {Silk}, J., \& {Schnittman}, J.~D. 2019, \apjl, 885, L35,
  1905.03790

\bibitem[{{Naoz} {et~al.}(2020){Naoz}, {Will}, {Ramirez-Ruiz}, {Hees}, {Ghez},
  \& {Do}}]{Naoz+20}
{Naoz}, S., {Will}, C.~M., {Ramirez-Ruiz}, E., {Hees}, A., {Ghez}, A.~M., \&
  {Do}, T. 2020, \apjl, 888, L8, 1912.04910

\bibitem[{{Pan} \& {Yang}(2021)}]{Pan+21}
{Pan}, Z., \& {Yang}, H. 2021, \prd, 103, 103018, 2101.09146

\bibitem[{{Pesce} {et~al.}(2018){Pesce}, {Braatz}, {Condon}, \&
  {Greene}}]{Pesce+18}
{Pesce}, D.~W., {Braatz}, J.~A., {Condon}, J.~J., \& {Greene}, J.~E. 2018,
  \apj, 863, 149, 1807.04598

\bibitem[{{Peters} \& {Mathews}(1963)}]{Peters+63}
{Peters}, P.~C., \& {Mathews}, J. 1963, Physical Review, 131, 435

\bibitem[{{Preto} \& {Amaro-Seoane}(2010)}]{Preto+10}
{Preto}, M., \& {Amaro-Seoane}, P. 2010, \apjl, 708, L42, 0910.3206

\bibitem[{{Rauch} \& {Tremaine}(1996)}]{Rauch+96}
{Rauch}, K.~P., \& {Tremaine}, S. 1996, NA, 1, 149, astro-ph/9603018

\bibitem[{{Raveh} \& {Perets}(2021)}]{Raveh+21}
{Raveh}, Y., \& {Perets}, H.~B. 2021, \mnras, 501, 5012, 2011.13952

\bibitem[{{Robertson} {et~al.}(2006){Robertson}, {Bullock}, {Cox}, {Di Matteo},
  {Hernquist}, {Springel}, \& {Yoshida}}]{Robertson+06}
{Robertson}, B., {Bullock}, J.~S., {Cox}, T.~J., {Di Matteo}, T., {Hernquist},
  L., {Springel}, V., \& {Yoshida}, N. 2006, \apj, 645, 986, astro-ph/0503369

\bibitem[{{Robson} {et~al.}(2018){Robson}, {Cornish}, \& {Liu}}]{Robson+18}
{Robson}, T., {Cornish}, N., \& {Liu}, C. 2018, arXiv e-prints, 1803.01944

\bibitem[{{Robson} {et~al.}(2019){Robson}, {Cornish}, \& {Liu}}]{Robson+19}
{Robson}, T., {Cornish}, N.~J., \& {Liu}, C. 2019, Classical and Quantum
  Gravity, 36, 105011, 1803.01944

\bibitem[{{Rodriguez} {et~al.}(2006){Rodriguez}, {Taylor}, {Zavala}, {Peck},
  {Pollack}, \& {Romani}}]{Rodriguez+06}
{Rodriguez}, C., {Taylor}, G.~B., {Zavala}, R.~T., {Peck}, A.~B., {Pollack},
  L.~K., \& {Romani}, R.~W. 2006, \apj, 646, 49, astro-ph/0604042

\bibitem[{{Rose} {et~al.}(2020){Rose}, {Naoz}, {Gautam}, {Ghez}, {Do}, {Chu},
  \& {Becklin}}]{Rose+20}
{Rose}, S.~C., {Naoz}, S., {Gautam}, A.~K., {Ghez}, A.~M., {Do}, T., {Chu}, D.,
  \& {Becklin}, E. 2020, \apj, 904, 113, 2008.06512

\bibitem[{{Rose} {et~al.}(2021){Rose}, {Naoz}, {Sari}, \& {Linial}}]{Rose+22}
{Rose}, S.~C., {Naoz}, S., {Sari}, R., \& {Linial}, I. 2021, arXiv e-prints,
  arXiv:2201.00022, 2201.00022

\bibitem[{{Rubbo} {et~al.}(2006){Rubbo}, {Holley-Bockelmann}, \&
  {Finn}}]{Rubbo+06}
{Rubbo}, L.~J., {Holley-Bockelmann}, K., \& {Finn}, L.~S. 2006, \apjl, 649, L25

\bibitem[{{Runnoe} {et~al.}(2017){Runnoe}, {Eracleous}, {Pennell}, {Mathes},
  {Boroson}, {Sigursson}, {Bogdanovc}, {Halpern}, {Liu}, \&
  {Brown}}]{Runnoe+17}
{Runnoe}, J.~C. {et~al.} 2017, MNRAS, 468, 1683, 1702.05465

\bibitem[{{Sari} \& {Fragione}(2019)}]{Sari+19}
{Sari}, R., \& {Fragione}, G. 2019, \apj, 885, 24, 1907.03312

\bibitem[{{Schnittman}(2015)}]{Schnittman15}
{Schnittman}, J.~D. 2015, \apj, 806, 264, 1506.06728

\bibitem[{{Schnittman} {et~al.}(2018){Schnittman}, {Dal Canton}, {Camp},
  {Tsang}, \& {Kelly}}]{Schnittman+18}
{Schnittman}, J.~D., {Dal Canton}, T., {Camp}, J., {Tsang}, D., \& {Kelly},
  B.~J. 2018, \apj, 853, 123, 1704.07886

\bibitem[{{Sillanpaa} {et~al.}(1988){Sillanpaa}, {Haarala}, {Valtonen},
  {Sundelius}, \& {Byrd}}]{Sillanpaa+88}
{Sillanpaa}, A., {Haarala}, S., {Valtonen}, M.~J., {Sundelius}, B., \& {Byrd},
  G.~G. 1988, \apj, 325, 628

\bibitem[{{Smith} {et~al.}(2010){Smith}, {Shields}, {Bonning}, {McMullen},
  {Rosario}, \& {Salviander}}]{Smith+10}
{Smith}, K.~L., {Shields}, G.~A., {Bonning}, E.~W., {McMullen}, C.~C.,
  {Rosario}, D.~J., \& {Salviander}, S. 2010, \apj, 716, 866, 0908.1998

\bibitem[{{Sridhar} \& {Touma}(2016)}]{Sridhar+16}
{Sridhar}, S., \& {Touma}, J.~R. 2016, \mnras, 458, 4143, 1509.02401

\bibitem[{{Stemo} {et~al.}(2020){Stemo}, {Comerford}, {Barrows}, {Stern},
  {Assef}, {Griffith}, \& {Schechter}}]{Stemo+20}
{Stemo}, A., {Comerford}, J.~M., {Barrows}, R.~S., {Stern}, D., {Assef}, R.~J.,
  {Griffith}, R.~L., \& {Schechter}, A. 2020, arXiv e-prints, arXiv:2011.10051,
  2011.10051

\bibitem[{{Touma} {et~al.}(2019){Touma}, {Tremaine}, \&
  {Kazandjian}}]{Touma+19}
{Touma}, J., {Tremaine}, S., \& {Kazandjian}, M. 2019, \prl, 123, 021103,
  1907.01555

\bibitem[{{Tremaine} {et~al.}(2002){Tremaine}, {Gebhardt}, {Bender}, {Bower},
  {Dressler}, {Faber}, {Filippenko}, {Green}, {Grillmair}, {Ho}, {Kormendy},
  {Lauer}, {Magorrian}, {Pinkney}, \& {Richstone}}]{Tremaine+02}
{Tremaine}, S. {et~al.} 2002, \apj, 574, 740, astro-ph/0203468

\bibitem[{{van den Bosch}(2016)}]{van_den_Bosch16}
{van den Bosch}, R. C.~E. 2016, \apj, 831, 134, 1606.01246

\bibitem[{{Will} \& {Maitra}(2017)}]{Will+17}
{Will}, C.~M., \& {Maitra}, M. 2017, \prd, 95, 064003, 1611.06931

\bibitem[{{Yunes} {et~al.}(2008){Yunes}, {Sopuerta}, {Rubbo}, \&
  {Holley-Bockelmann}}]{Yunes+08}
{Yunes}, N., {Sopuerta}, C.~F., {Rubbo}, L.~J., \& {Holley-Bockelmann}, K.
  2008, \apj, 675, 604, 0704.2612

\bibitem[{{Zheng} {et~al.}(2020){Zheng}, {Lin}, \& {Mao}}]{Zheng+20}
{Zheng}, X., {Lin}, D. N.~C., \& {Mao}, S. 2020, arXiv e-prints,
  arXiv:2011.04653, 2011.04653

\end{thebibliography}
\end{document}